\documentclass[smallextended]{./LaTeX_DL_468198_240419/svjour3}       
\smartqed  
\usepackage{graphicx}
\usepackage{natbib}

\usepackage[%
  pdfstartview=FitH,%
  bookmarks=false,%
  bookmarksopen=false,%
  breaklinks=true,%
  colorlinks=true,%
  linkcolor=blue,anchorcolor=blue,%
  citecolor=blue,filecolor=blue,%
  menucolor=blue,pagecolor=blue,%
  urlcolor=blue]{hyperref}
  
\usepackage{url}

\usepackage{graphics,appendix,rotating,setspace}
\usepackage{amsfonts, amssymb,amsmath,amsbsy}
\usepackage[english]{babel}
\usepackage[latin1]{inputenc}
\usepackage[T1]{fontenc}

\usepackage{fullpage}
\usepackage{lineno}
\usepackage{wrapfig}
\usepackage{subfig}
\usepackage[outdir=./Figures/]{epstopdf}

\usepackage{soul}
\usepackage{ulem}
\usepackage{color}

\begin{document}

\title{Rapid variations of Earth's core magnetic field
}
\subtitle{{\tt Version from the:} \today}

\titlerunning{Core field fast variations}        

\author{V. Lesur \and N. Gillet \and M.D.  Hammer \and M. Mandea}

\authorrunning{Lesur et al.} 

\institute{V. Lesur \at
             Université de Paris, Institut de physique du globe de Paris, CNRS, Paris, France\\
             \email{lesur@ipgpg.fr}  
             \and
             N. Gillet \at 
             Univ. Grenoble Alpes, Univ. Savoie Mont Blanc, CNRS, IRD, UGE, ISTerre, 38000 Grenoble, France
             \and
             M.D. Hammer \at
             Division of Geomagnetism and Geospace, DTU Space, Technical University of Denmark 
            \and
             M. Mandea \at
             CNES -- Centre National d'Etudes Spatiales, 2 place Maurice Quentin 75039 Paris Cedex 01}

\date{Received: date / Accepted: date}

\maketitle

\begin{abstract}
Evidences of fast variations of the Earth's core field are seen both in magnetic observatory and satellite records. We present here how they have been identified at the Earth' surface from ground-based observatory records 
and how their spatio-temporal structure is now characterised by satellite data. It is shown how their properties at the core mantle boundary are extracted through localised and global modelling processes, paying particular attention to their time scales. Finally are listed possible types of waves in the liquid outer core, together with their main properties, that may give rise to these observed fast variations. 


\end{abstract}

\section{Introduction}
\label{sec: intro}

The observed magnetic field above the Earth's surface results from the contribution of numerous sources situated inside the Earth, such as the core and the lithosphere, or outside the Earth, such as the ionosphere and magnetosphere. 
The dominant contribution is by far the field generated in the core. 
Its longer time-changes, including millennial and longer periods, are 
reconstructed from paleomagnetism \citep{constable2005paleomagnetic}. 
The shortest fluctuations likely reach sub-daily periods, in link with the rotating fluid dynamics of the fluid core.  
However, the variations directly observable at the Earth' surface are on a limited range of time-scales. 
The upper limit is slightly longer than a century, corresponding to the installation of the first magnetic observatories \citep[e.g.,][]{matzka2010geomagnetic}. 
The lower limit is poorly known for two reasons.
Firstly, the amplitude of the core signal gets weaker towards high frequencies, so that external signals dominate the magnetic records on periods shorter than $O(1)$ year. Secondly, the screening effect of the conductive mantle, which filters out rapid signals originating from the core, is not well estimated. 
The aim of this paper is to review our current knowledge on the rapid variations of the core field, and the possible core flow perturbations that generate them. \\

\noindent Very early in the study of the Earth's magnetism it had been realised that the magnetic field ${\bf B}$ is changing with time \citep[e.g.,][]{kono2007geomagnetism}. 
The proper description of this evolution came with the setting of magnetic observatories from the middle of the 19th century onward. 
The general assumption was that most of observed fast variations in ${\bf B}(t)$ were due to external fields perturbations,
while the dominant core field varied slowly over time. 
The accumulation of long time series of magnetic observations led to an evolution of this paradigm since, at observatory sites, the main field secular variation (SV, or rate of change of the field, $\partial{\bf B}/\partial t$) often appeared as linear trends with abrupt changes of slopes. 
These singular events are clearly generated in the core and have been called "geomagnetic jerks" \citep{Courtillot:1978}. 
{Early discussions on the origin of these events can be found in} \cite{Malin:1982,  Alldredge:1984}. 
The evolution of this secular variation has also been studied in terms of secular acceleration (SA, or second time derivative, $\partial^2{\bf B}/\partial t^2$). 
Early models of the SA can be found in \cite{Cain:1967}, who used OGO satellite data, assuming the acceleration constant over time. 
The acceleration was also assumed constant by \cite{Malin:1969} and \cite{Barraclough:1979}, although in these latter publications it is already noticed that the difference of SA obtained by different authors likely reveal a temporal change. 
After the \textsc{magsat} satellite mission in 1980, came the first robust representations of magnetospheric fields \citep{Langel:1985a, Langel:1985b}, as well as temporal parameterisations of the core field changes using \textsc{b}-splines \citep{Langel:1986, Bloxham:1992, Sabaka:1997}. 
In principle such models are able to describe the SA and its evolution, but these are not discussed in the aforementioned publications.
After the launch of the Oersted and \textsc{champ} satellite missions, it became possible to derive more accurate magnetic field models. 
While preparing candidate models for the 10th \textsc{igrf} generation \citep[International Geomagnetic Reference Field, ][] {IGRF:2005}, it became obvious that assuming a linear evolution of the magnetic field was not adequate to fit the few years of available satellite data: including an acceleration was necessary \citep{Lesur:2005, Olsen:2005, Maus:2005a}. 
Soon after came models fitting these data and with a temporal description of the core field using \textsc{b}-splines, initially of order four as in the first versions of the \textsc{chaos} model \citep{Olsen:2006, Olsen:2008}, and then five in the \textsc{grimm} model \citep{Lesur:2008}. 
With this latter model it was possible to follow the SA variations over five years, and to identify a maximum of acceleration in 2006. 
From there on, numerous publications \citep[e.g.,][]{chulliat2014geomagnetic,Finlay_etal_2016a} have focused on the acceleration patterns, highlighting pulses in the SA norm at inter-annual periods. These are discussed in the following of this paper.\\

\noindent The crucial issue regarding field models derived from magnetic data is their temporal resolution.
Observatory time series show a temporal spectrum $S(f)\propto f^{\alpha}$ with a slope $\alpha\simeq -4$ in the range of periods from $\approx 70$ yrs down to a couple of years \citep{de2003spatial}.  
This property, which is coherent with the existence of geomagnetic jerks, is also recovered in time series of the geomagnetic Gauss coefficients \citep{lesur2018frequency}. 
It is shared by the family of stochastic processes (of order 2) that is differentiable only once in time \citep{Gillet:2013}. 
For those, the estimate of the SA is subject to the choice of sampling rate. 
With most field models, an instantaneous measure of the SA is not available, but instead only a filtered vision. 
This time resolution strongly depends on the spherical harmonic degree: the larger the length-scale, the better the resolution in frequency (see section~\S\ref{sec:gfm}). 
A magnetic signal probably exist at inter-annual periods, which carries some information about the core physics. 
However, the image of the shorter wavelengths is limited to long periods, which alters our reconstruction of inter-annual SA patterns. 
This limitation precludes the identification of the cutting frequency over which the slope $S(f)\propto f^{-4}$ stops.
This depends on the one hand on the conductance of the mantle \citep{jault2015conductivity}, and on the other hand on the relative importance of diffusion and propagation of Alfv\'en waves in the fluid outer core \citep{aubert2021}.\\

\noindent There is a general agreement that the flow pattern in the liquid outer core that generates most of the observed SV (by advection of the core field), is a large gyre that flows westward near the core mantle boundary under Africa and the Atlantic, and flows closer to the cylinder aligned with the rotation axis and tangent to the inner core 
under the Pacific Ocean \citep[e.g.,][]{Pais:2008, gillet2009ensemble, gillet2015planetary, aubert2014earth, barenzung2018modeling}. 
It has been demonstrated however, that this flow has to vary in time to explain the observed magnetic field variations at observatory sites \citep{bloxham1992steady,Waddington:1995}. 
There are some indications that these flow variations are mainly located under Eastern Asia, and Central America \citep[e.g.,][]{Finlay_etal_2016a}, but they have also been identified close to the inner core tangent cylinder \citep{Livermore:2017,gillet2019reduced}. 
Torsional waves with a periodicity around 6~years have also been identified in flow models derived from ground based observations \citep{Gillet:2010N}. 
They explain an independent 6-yr signal recorded in the length-of-day \citep{abarca2000interannual,holme2013characterization,chao2014earth}, and their physics determine the intensity of the magnetic field within the core (where the potential field approximation drops due to electrical currents).  
The magnetic signature of these waves at Earth's surface is tiny (a couple of nT.yr$^{-1}$ --  e.g., \cite{Cox:2016}), and must be coherent within the set of observation sites in order to imprint core flow models. 
This magnitude is nevertheless too weak to entirely explain inter-annual SV changes (of the order of $8$ nT.yr$^{-1}$), which thus require the existence of other, non-zonal, transient motions \citep{gillet2015planetary,Finlay_etal_2016a,Kloss_Finlay_2019}.\\

\noindent The remaining of this paper is organised as follow: in section \S\ref{sec: obs} we first come back on long series of observatory data and subsequently satellite (GVO) data, from which we discuss the observed SV evolutions over time. 
We present next in section \S\ref{sec: modeling} different modelling techniques and evaluate how these affect the derived acceleration time scales. 
Then, possible flow structures and waves that may generate observable signals at Earth surface are reviewed in section \S\ref{sec: core physics}. 
We finally conclude in section \S\ref{sec: conclusion}.

\section{Core field variability observed from magnetic data}
\label{sec: obs}

\subsection{Ground magnetic observatories}
\label{sec: GO}

\begin{figure}
    \centering
 \includegraphics[width=10. cm, keepaspectratio=true]{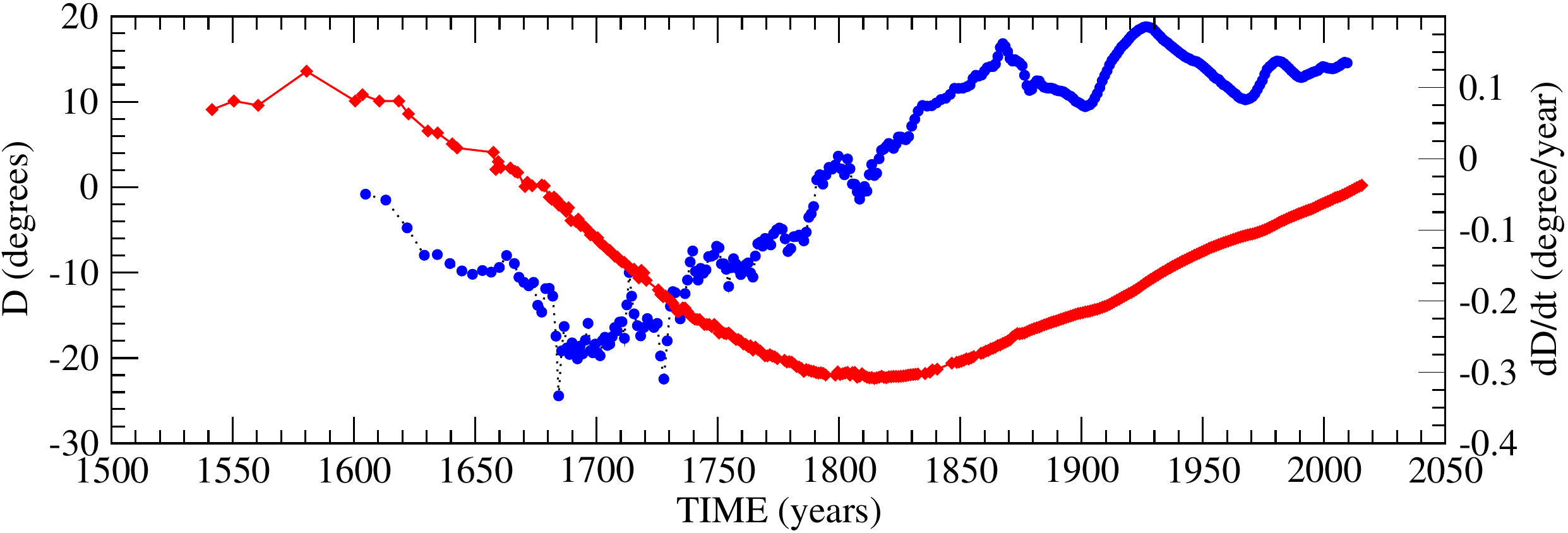}
    \caption{Compiled records of the observed declination ($D$, in red) and its estimated secular variation (SV, in blue), in the vicinity of Paris, France. All records have been reduced to a single position at the current site of the French national magnetic observatory in Chambon-la-Forêt (\textsc{clf} -- Loiret, France). }
    \label{fig:Paris_D}
\end{figure}

A strong motivation in running magnetic observatories is 
the monitoring of the Earth's main magnetic field evolution originating in the fluid outer core. The importance of long-term observations in support of this endeavour is illustrated in Figure~(\ref{fig:Paris_D}) where (irregular) declination measurements at a number of locations in the vicinity of Paris, and regular measurements at the observatories in Saint Maur des Fossés (Val de Marne, France, 1883-1902), Val Joyeux (Yvelines, France, 1903-1935) and Chambon-la-For\^et (1936-present) are plotted together. The Paris declination series starts as early as in the 16th century, with the first measurement performed in 1541, giving a value of $7^{\circ} E$. Let us underline that all declination measurements made in the Paris area (Issy-les-Moulineaux, Montsouris et Montmorency) and in previous French magnetic observatories are reduced to the current location of Chambon-la-Forêt. A detailed discussion of the applied corrections is given in \cite{alexandrescu_paris_1996}. The plot shows that, over the last 479 years, the magnetic declination has varied between extreme values reaching $13^{\circ} 30' E$ in 1580 to $22^{\circ}30' W$ in 1814. Long data series, such as those in Figure~(\ref{fig:Paris_D}), have been compiled for example for London, Rome and Munich by \cite{malin_direction_1981, cafarella_secular_1992, korte_historical_2009}, respectively. These series are important in characterising the scales of temporal variations in the core-generated field. \\

\noindent These unique series allow to easily compute the SV in a single position and to estimate occurrence date for geomagnetic jerks, mainly before the 20th century. In Figure~(\ref{fig:Paris_D}) the SV is computed as annual mean differences. To enhance rapid events and reduce solar cycle effects, the first time derivative is computed after applying an 11-yr smoothing low-pass filter window which makes the changes in the trend more rounded. Nevertheless, the geomagnetic jerks can still be clearly identified. The figure clearly shows that, prior to the 20th century, one of the most prominent geomagnetic jerks appears around 1870. This event is also observed in four other European locations \citep{alexandrescu_high-resolution_1997} and has recently been detected in the Munich curve \citep{korte_historical_2009}, although a few years earlier. 
Going farther back in time, there is evidence of changes in the secular variation trend, supported by measurements around  epochs 1600, 1665, 1700, 1730, 1750, 1760, 1770, 1810, 1870, 1890, and 1900. These dates are close to those detected by \cite{qamili_geomagnetic_2013}: 1603, 1663, 1703, 1733, 1751, 1763, 1770, 1810, 1868, 1870, 1888, and 1900, when analysing the temporal behaviour of the difference between predicted and actual geomagnetic field model values for successive intervals from 1600 to 1980, based on the \textsc{gufm}1 geomagnetic model \citep{jackson2000four}. \\

\noindent The typical SV time scale spans years to centuries, and its typical magnitude is $10-100$ nT.yr$^{-1}$. These values help to define the observational requirements for magnetic observatories: they should run for many years, achieve measurement accuracy in all components of the field vector of around 1 nT, and maintain the long-term stability needed to resolve the typical secular change signal. With the above recommendation the geomagnetic observatories represent a unique dataset to understand the secular variation over long-term periods. Indeed, the first geomagnetic jerk has been detected was identified in geomagnetic observatories series \citep{Courtillot:1978}. In this pioneering paper, it was defined as an abrupt turning point separating the otherwise linear trends of the East component of SV prior to and after 1970 at several northern hemisphere observatories. \\

\begin{figure}
    \centering
   \includegraphics[width=10cm, keepaspectratio=true]{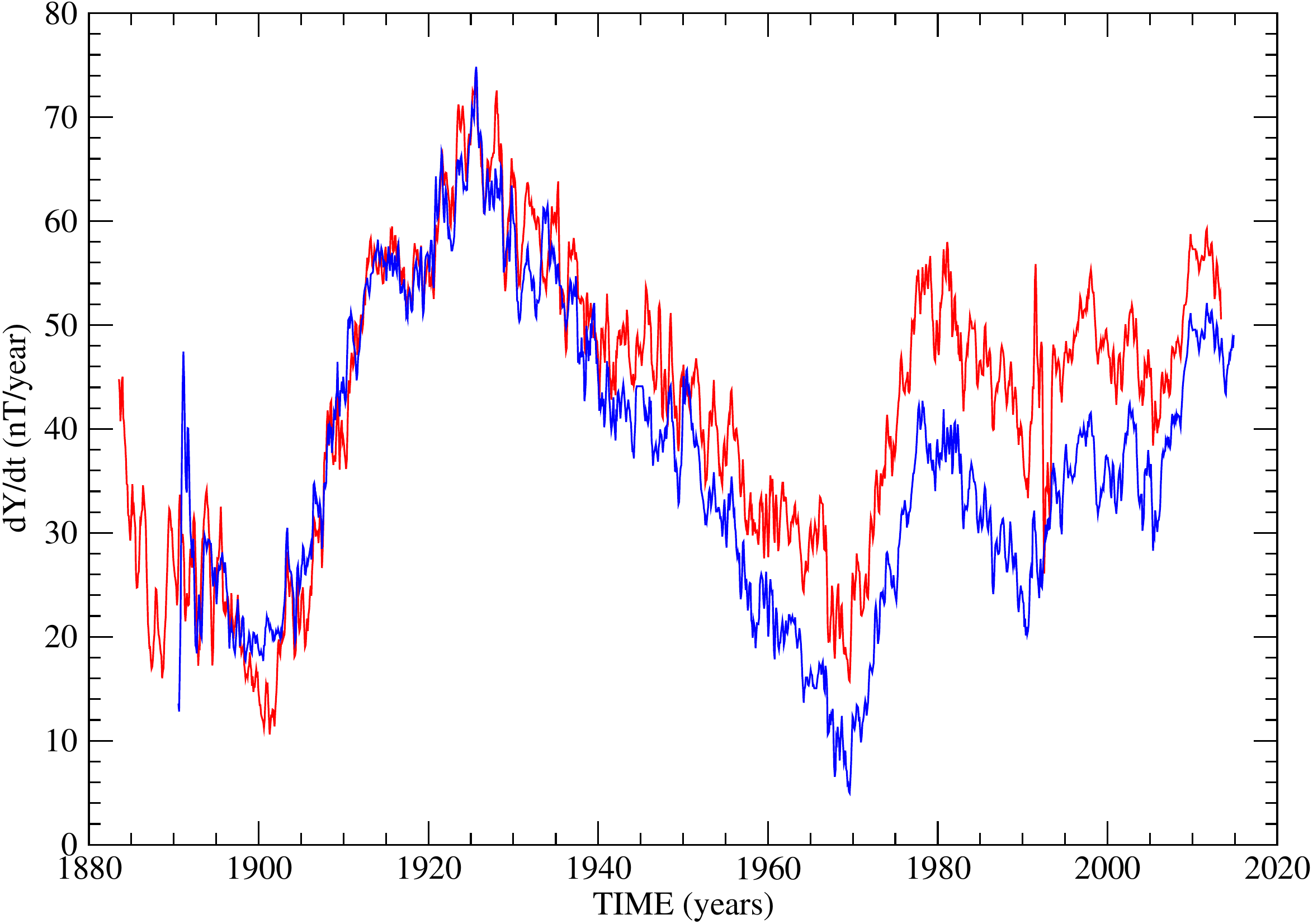}
    \caption{In red is shown the secular variation for the East component at Chambon-la-Forêt (\textsc{clf} -- 48.025$^\circ$N, 2.260$^\circ$E), France and, in blue, at Niemegk (\textsc{ngk} -- 52.07$^\circ$N, 12.68$^\circ$E), Germany.  }
    \label{fig:clf_ngk}
\end{figure}

\noindent The geomagnetic jerks can be observed in the various field derivatives
(core field, secular variation and secular acceleration) with a wide variety of detection methods. A detection method needs to consider several factors linked to the available data: noise content in the data and its origin, the asynchronous form of a jerk in each field component, temporal and spatial scales at which an event is significant enough to be seen as a jerk. These factors are briefly discussed here:
\begin{itemize}
\item[-] the quality of data varies back in time. Systematic geomagnetic field observations at multiple locations, for the three field components, began with the establishment, by Gauss, of the G{\"o}ttingen Magnetic Union in 1834. Before that date only directional observations were available, so the characterisation of the secular variation via the full vector field was not possible. Only over the 20th century the three field components are available for a set of observatories allowing comparison between different locations and different field components.
\item[-]  the first data analyses were focused on the East component, mainly for two reasons. The first analyses considered mainly European observatories and the geomagnetic jerks are generally clearly noted in this component time series (see Figure~\ref{fig:clf_ngk}). It should be noted that the behaviour of the East component secular variation is different for \textsc{clf} (48.025$^\circ$N, 2.260$^\circ$E) and \textsc{ngk} (52.07$^\circ$N, 12.68$^\circ$E) observatories, even if these two observatories are relatively close.
\item[-]  the magnetic field Eastern component is less affected by the external sources, i.e. electrical currents flowing in the ionosphere and magnetosphere of the Earth. Different methods have been applied to minimise the external field contributions in geomagnetic series (monthly or annual means), via geomagnetic indices or a combination of field models and magnetic indices, or empirical techniques \citep[e.g.,][]{de_michelis_global_2000, verbanac_contributions_2007,wardinski2011signal}. Different attempts have been also applied to parameterise the external field sources during the modelling processes (see section~\S\ref{sec:gfm} and references there in). These models are also suitable to obtain series of data and to investigate geomagnetic jerks, mainly when considering three field components.
\item[-]  it is well-known that the geomagnetic observatories are located mainly on continental areas and that the northern hemisphere is better covered than the southern hemisphere. It is then important to recall the effect of the uneven distribution of observatories in the definition of geomagnetic jerk characteristics. Identifying regional jerks have been possible not only because more and more sophisticated methods have been applied \citep[see][]{Mandea_etal_2010, Brown:2013}, but also due to the usage of full field geometry.
\end{itemize}

\noindent Although geomagnetic jerks are clearly of internal origin and are well identified events in magnetic observatories time series, they are less and less seen as singular events in term of flow dynamics. Study of recent magnetic field models and progresses in numerical geodynamo models suggest that they are natural features of the magnetic field generation in the Earth's core (see Section \S\ref{sec: core physics}). \\

\begin{figure}
{\center
\includegraphics[width=5cm, angle=0,keepaspectratio=true]{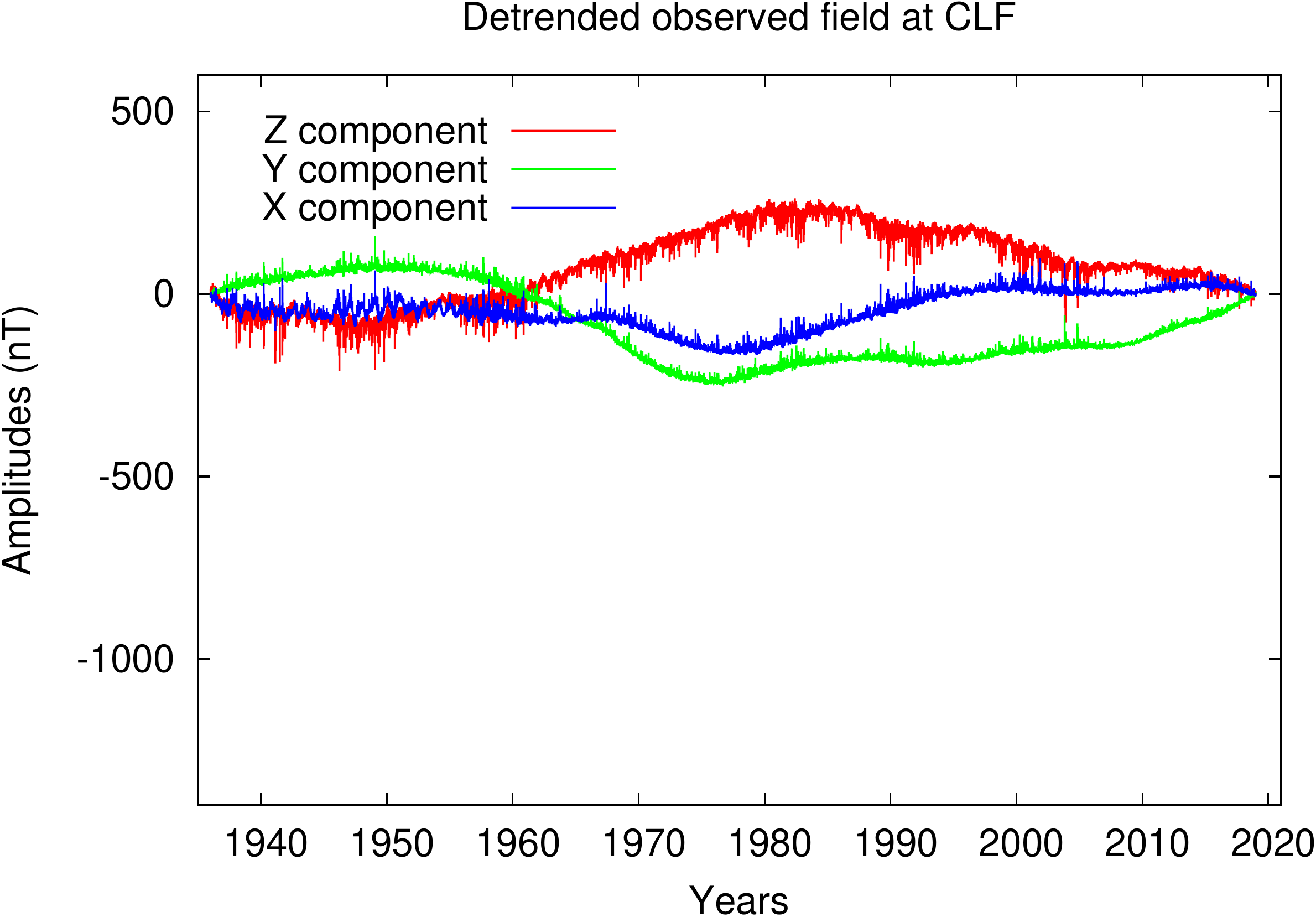}
\includegraphics[width=5cm, angle=0,keepaspectratio=true]{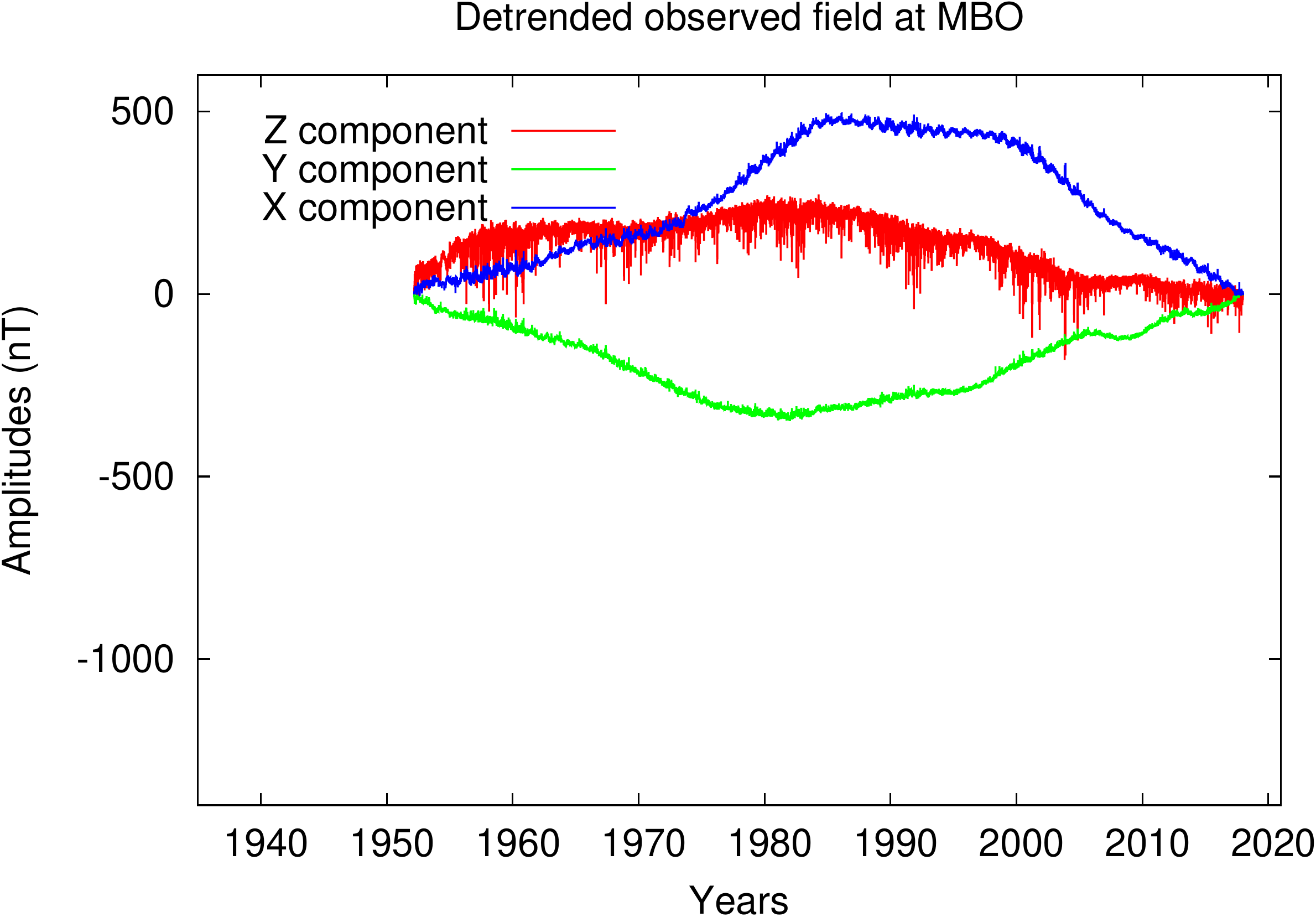}
\includegraphics[width=5cm, angle=0,keepaspectratio=true]{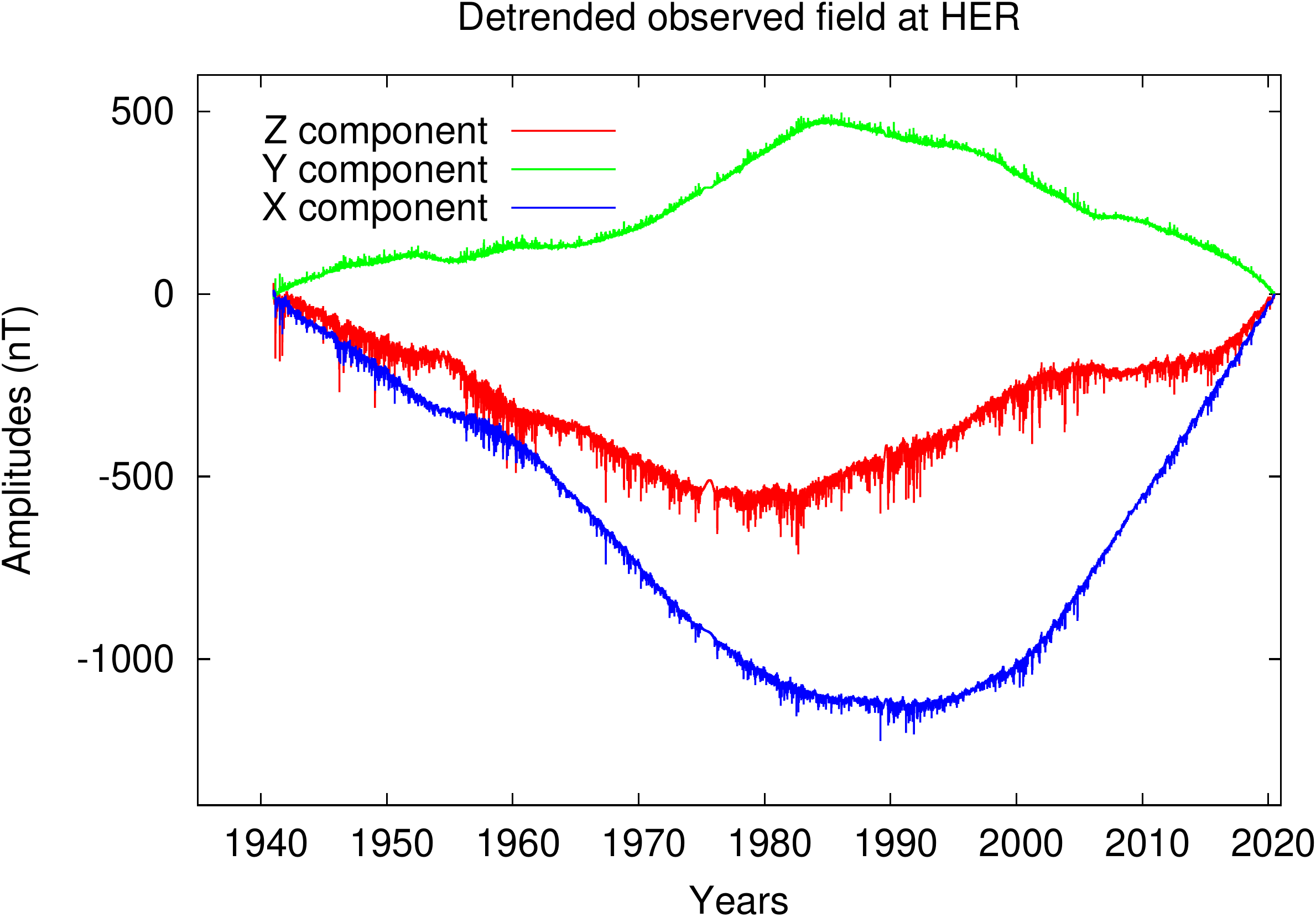} \vspace{0.3cm}\\
\includegraphics[width=5cm, angle=0,keepaspectratio=true]{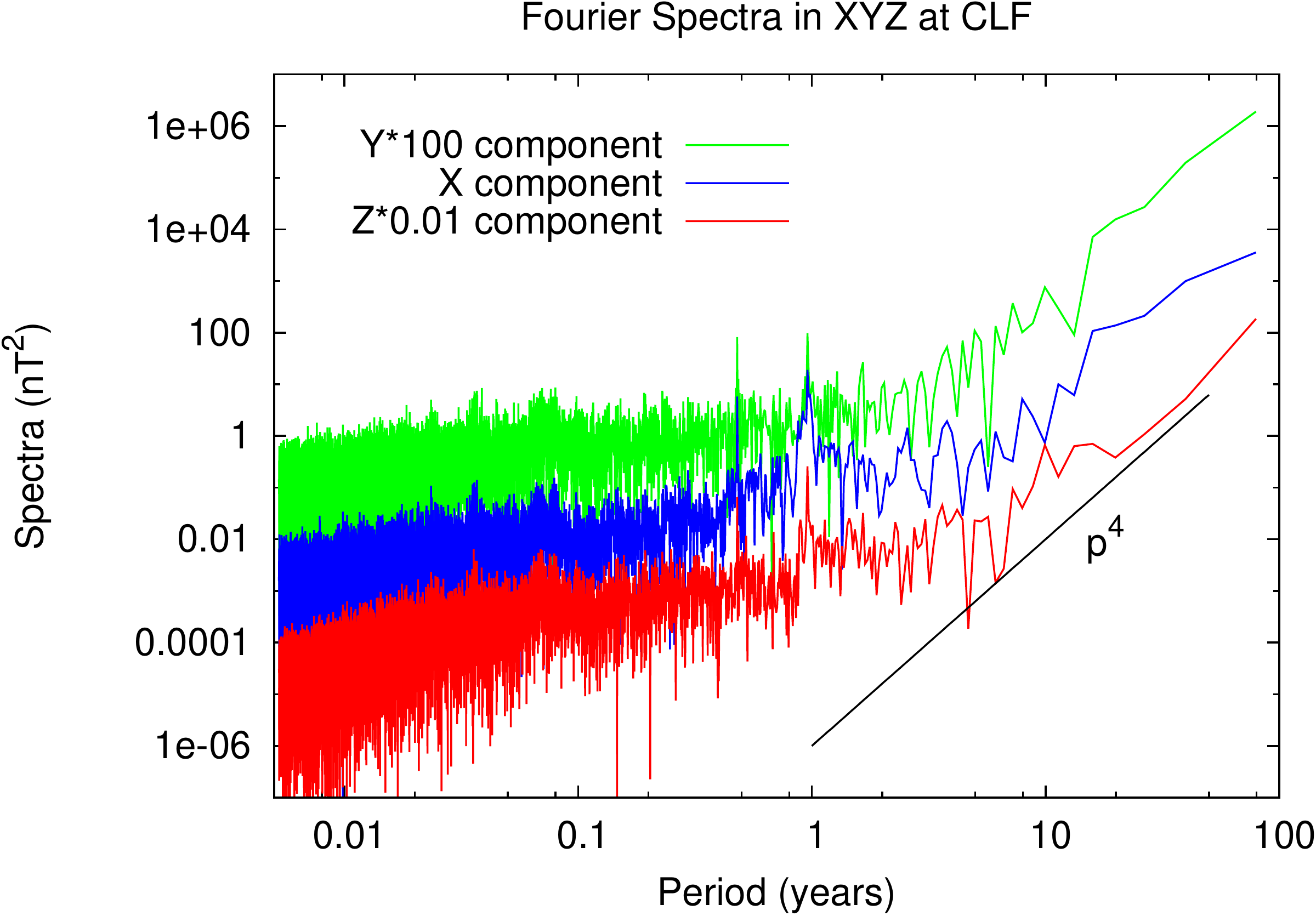}
\includegraphics[width=5cm, angle=0,keepaspectratio=true]{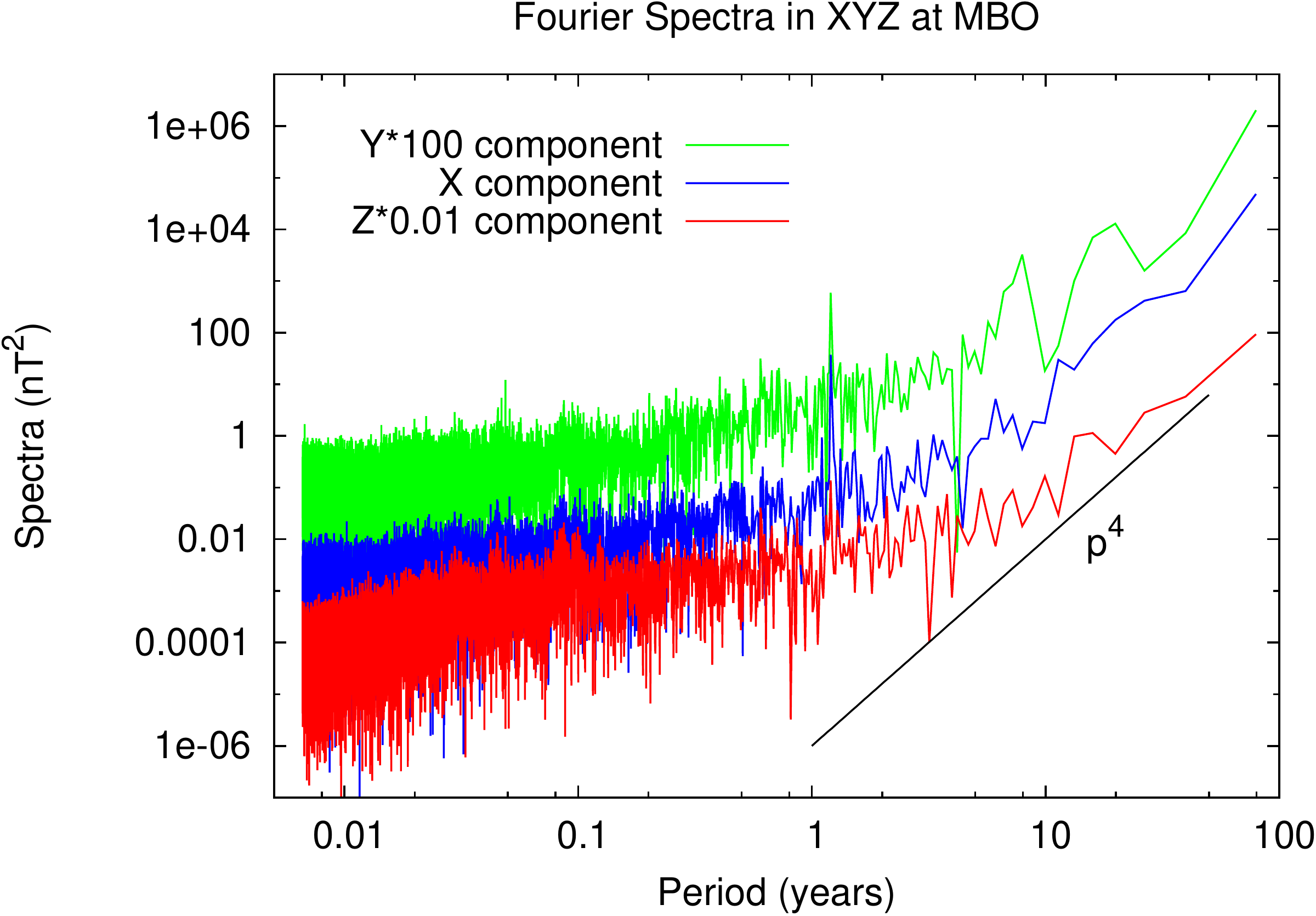}
\includegraphics[width=5cm, angle=0,keepaspectratio=true]{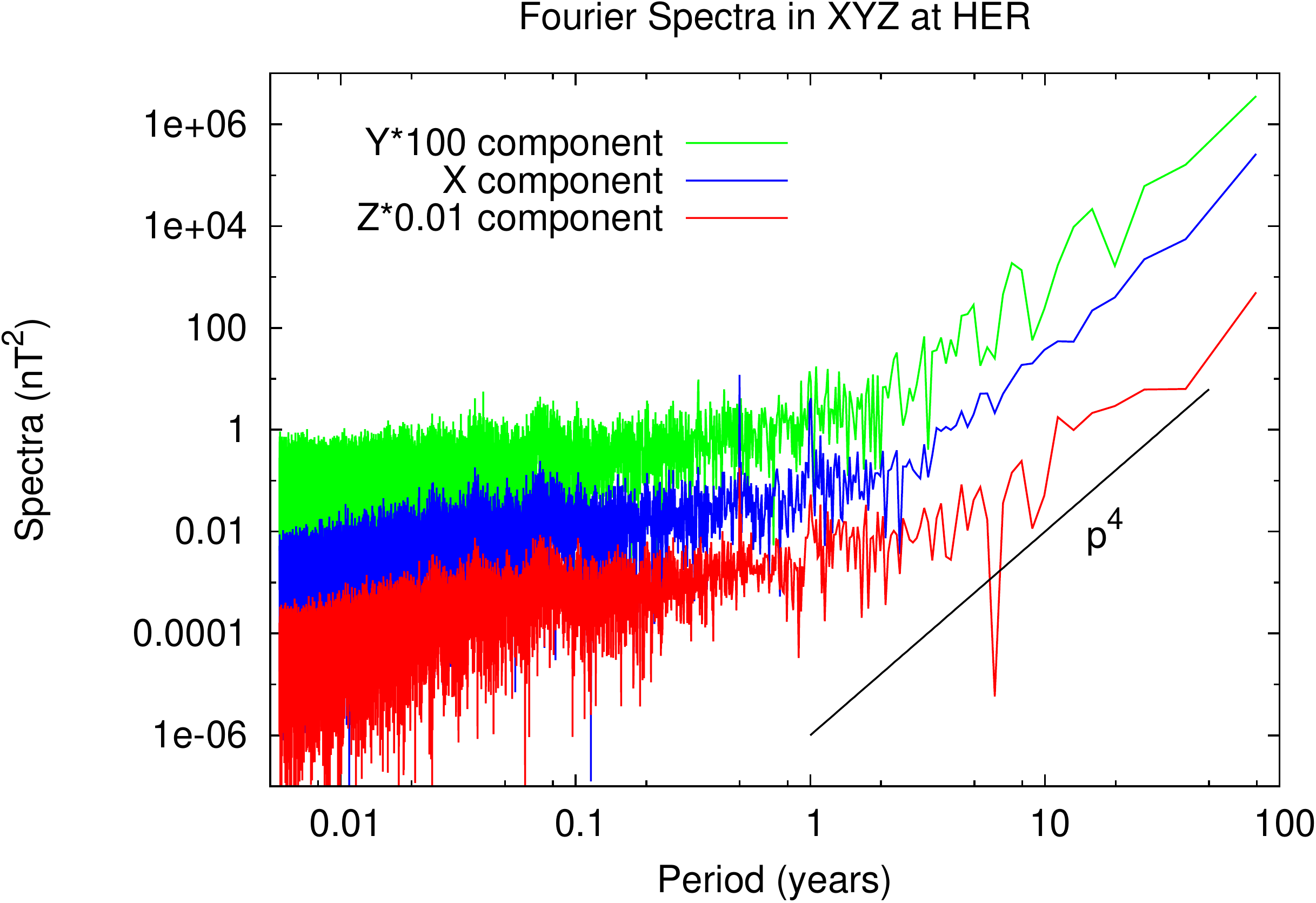}
\caption{First row: Detrended three components of the magnetic field recorded at observatory sites in \textsc{ECEF} system of coordinates. Second row: Fourier spectra of the corresponding time series where $Y$ and $Z$ components have been shifted up and down, respectively, for clarity. The three presented observatories are from left to right: \textsc{ clf, mbo, her} (see text for details). The slope $\propto p^4$, with $p$ the period, is indicated with a black line.}
\label{fig:FFT}
}
\end{figure}

\noindent In order to get a better understanding of the recorded signal variability at magnetic observatories, in Figure~(\ref{fig:FFT}) are presented the time series of the three components of the magnetic field in an Earth Centred Earth Fixed system of coordinates (\textsc{ECEF} --i.e. Z is the Earth's rotation axis, X points towards Greenwich meridian and  Y complete this system), together with their Fourier spectra. Data are daily means. The time series have been corrected for recorded jumps, interpolated for missing values using smoothed cubic-splines and "detrended" by subtracting a linear polynomial going through the first and last point of the series. Figure~(\ref{fig:FFT}) thus display deviations from a linear evolution of the recorded data. All plots have the same scales, and deviations are much larger in Hermanus observatory (\textsc{her}, -34.425$^\circ$N, 19.225$^\circ$E, South-Africa) than on M'bour and Chambon-la-Forêt observatories, (\textsc{mbo}, 14.392$^\circ$N, -16.958$^\circ$E, Senegal) and (\textsc{clf}, 48.025$^\circ$N, 2.260$^\circ$E, France), respectively. Because of the choice of coordinate system a large part of the signals generated in the 
magnetosphere is concentrated in the Z component. 
These signals are associated in the Fourier spectra with peaks at one year and half-year periods. They are also related to excess of energy around 27-day periods and other sub-harmonics. The 11-yr solar cycle generates small peaks in Z components at all observatories. However, it is extremely difficult to identify specific periodicities for signals generated in the core, as the external field contributions dominate the spectra up to periods as large as 2 to 5 years, depending on locations and components. There is a clear excess in energy at M'bour in the Y component for periods centred on 6.5 years, and also for periods from 13 to 15 years. Less distinctive are these same periods for Hermanus observatory.  However, at Chambon-la-Forêt, in spite of a very long series starting the 1st January 1936, these periods do not dominate the spectra. The only unexpected maximum is at 1.7 years periods in the Y component. All observatories at the longest periods present an increasing trend that, as expected, roughly correspond to a $p^4$ behaviour, with $p$ the period. However, these observations indicate clearly the difference in the content of the observatory records depending on their regions. 
\noindent 

\subsection{Virtual observatories}
\label{sec:GVO}

With the advent of magnetic field measurements from low Earth orbiting satellites, smaller spatial structures of the core field changes can be mapped due to almost complete global coverage. 
The last twenty years of continuous space records of the magnetic field, provide valuable insights into the field and how it is changing in space and time. 
In particular, magnetic field measurements from the Danish Oersted satellite (1999-2014), the German \textsc{champ} satellite (2000-2010)  and the European { Swarm} satellite trio mission (2013-) have dramatically increased the constraint from geomagnetic data \citep{Olsen_Stolle_2012, domingos2019temporal}. 
In addition to these, calibrated CryoSat-2 satellite measurements (2010-) can be used to generate further constraints on the field variations, in particular during the temporal gap 2010--2014 between \textsc{champ} and Swarm \citep{Olsen_etal_2020}. 
Although the physical mechanisms responsible for most of the field changes take place over time scales longer than the satellite era (the past 20 years), shorter inter-annual fluctuations may be characterised with satellite data alone. \\

\begin{figure*}[!htbp]
\centerline{
\includegraphics[width=0.3\textwidth, keepaspectratio=true]{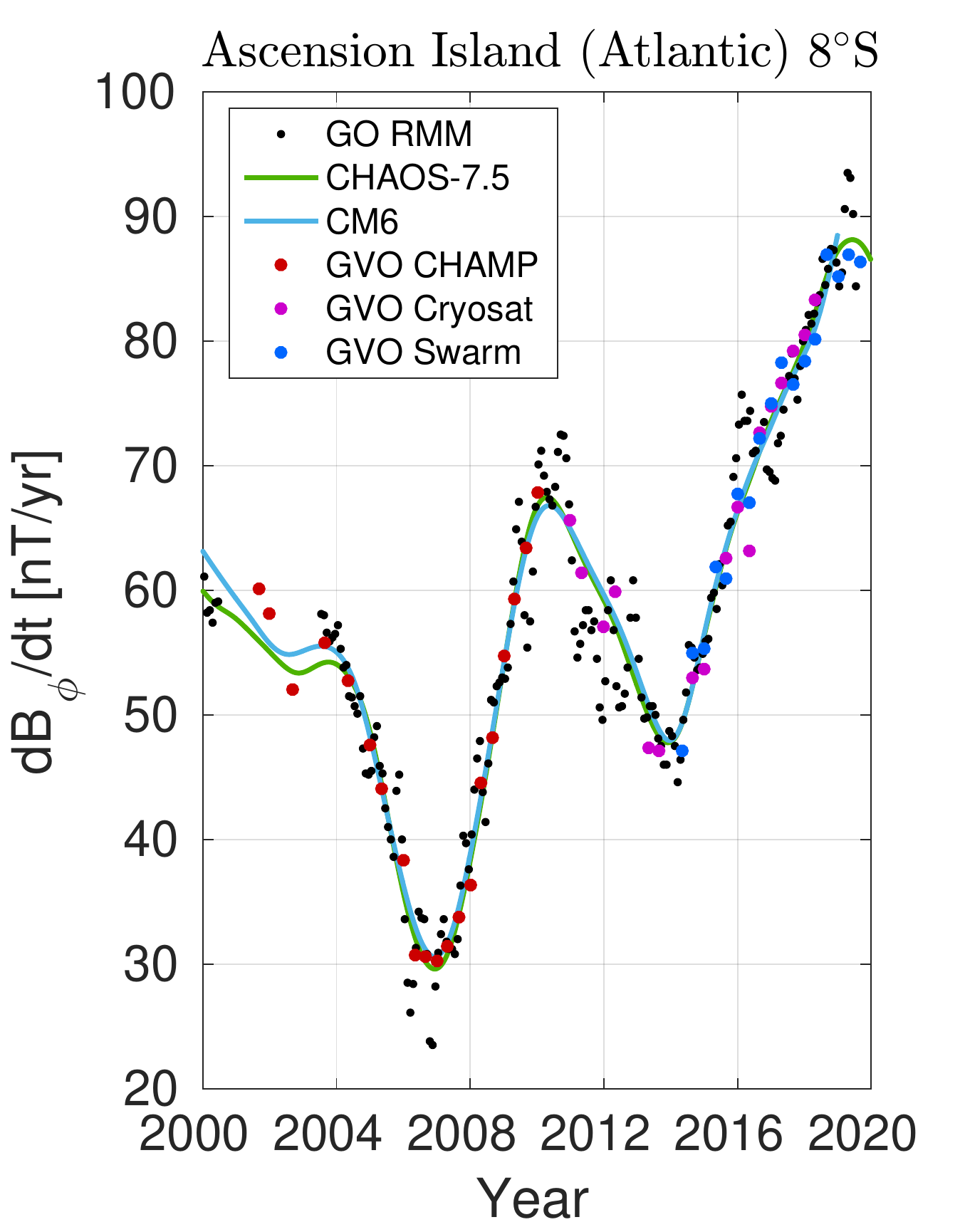}
\includegraphics[width=0.3\textwidth, keepaspectratio=true]{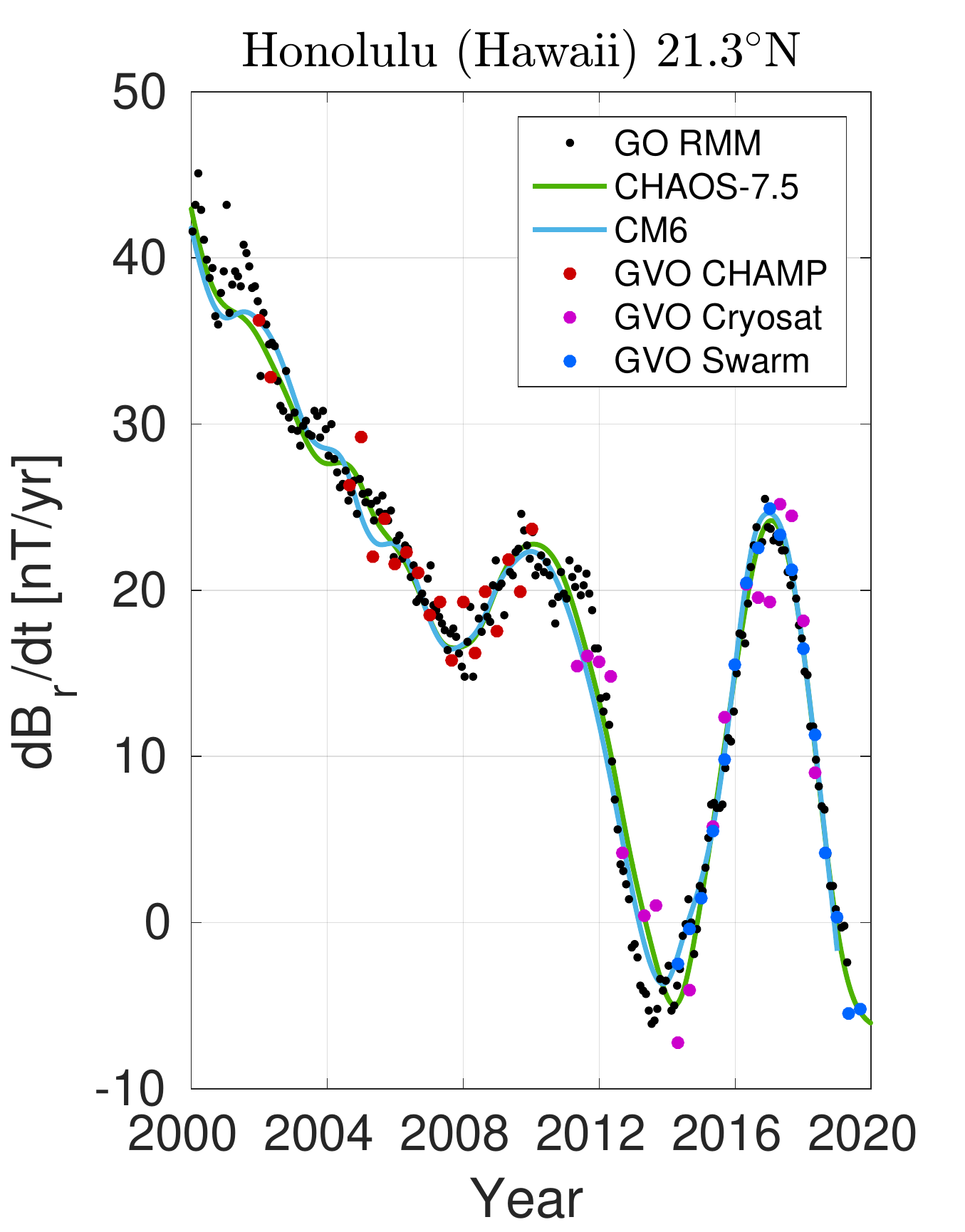}
\includegraphics[width=0.3\textwidth, keepaspectratio=true]{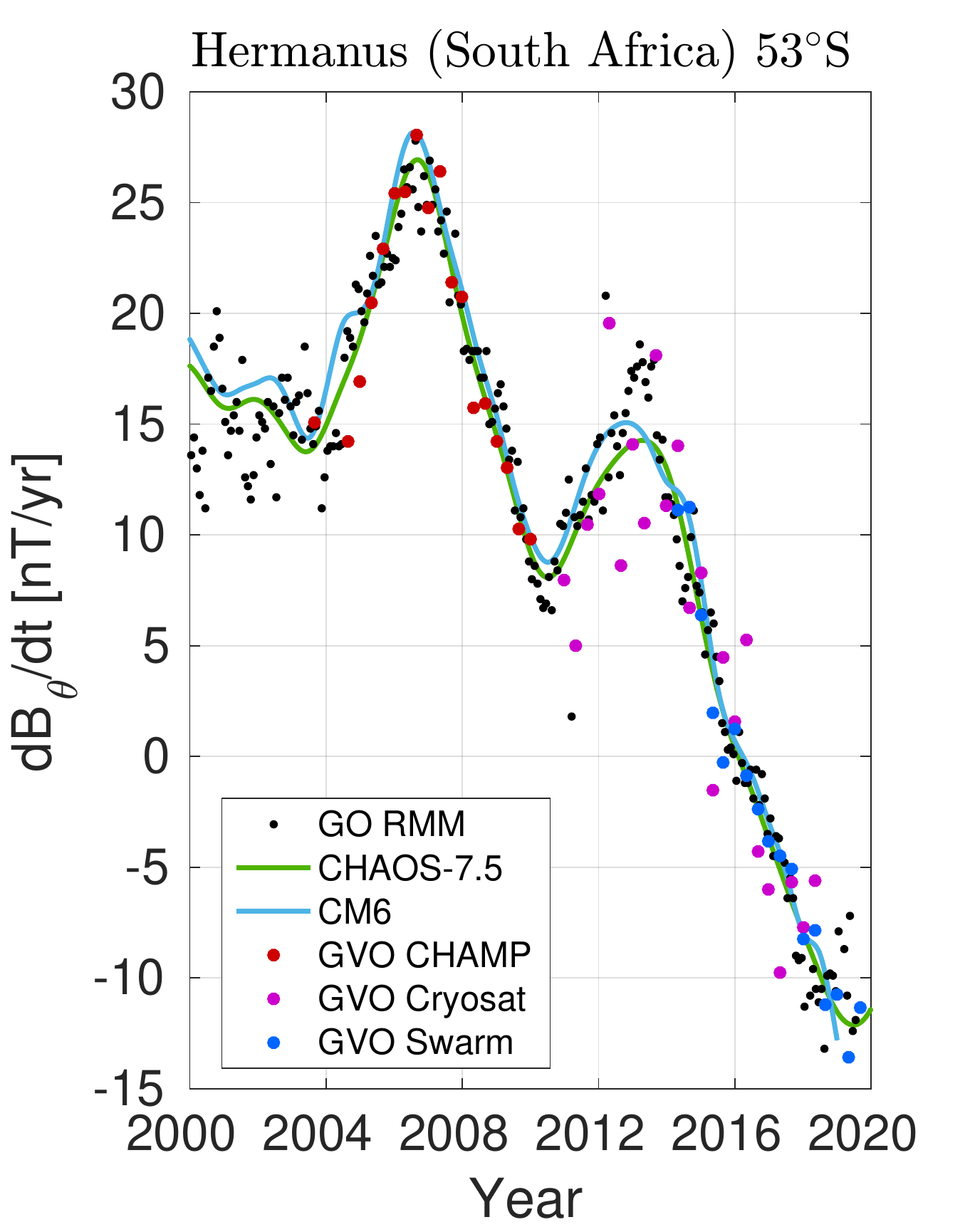}
}
\caption{Comparison of SV times series at three example ground observatories for the East ($\phi$), Radial ($r$), and South ($\theta$) components respectively. Shown are SV from annual differences of revised monthly means (black dots) \citep[see][]{olsen2014chaos}, and 4-monthly \textsc{gvo}s (mapped to ground) derived from \textsc{champ} (red dots), CryoSat-2 (purple dots) and {Swarm} (blue dots) measurements. Also shown are predictions of the CM6 \citep{Sabaka:2020} (cyan line) and \textsc{chaos}-7.5 \citep{Finlay:2020} (green line) models. 
}
\label{Fig: SV GVO comparison}
\end{figure*}

\noindent The Geomagnetic Virtual Observatory (\textsc{gvo}) technique was first developed by \cite{Mandea_Olsen_2006,Olsen_Mandea_2007}, as a way of making satellite field measurements easily accessible, through a regression of these into local time series of the field at pre-specified locations. 
Such \textsc{gvo} time series consist of vector field estimates at mean orbital altitude. 
Each \textsc{gvo} datum is derived using satellite measurements in the near vicinity of a given \textsc{gvo} location, and within a chosen time window. 
As for ground-based records, by taking annual differences of these locally derived \textsc{gvo} field estimates, SV time series can be obtained. 
To address issues of insufficient local time sampling and contamination from external sources such as the magnetospheric ring current and ionospheric current systems \citep{Olsen_Mandea_2007,Beggan_etal_2009,Shore_2013}, additional data processing steps, changes to the modelling scheme and de-noising procedures have been incorporated into the \textsc{gvo} algorithm \citep{Hammer_etal_2020a}. 
In order to make robust \textsc{gvo} field estimates, a pre-whitening step is applied wherein main field predictions ${\bf B}({\bf r}_{GVO})$ for each data point of position ${\bf r}_{GVO}$ are subtracted off, and at a later stage added back for the \textsc{gvo} epoch under consideration. 
The IGRF model (which has linear SV) is typically used for determining ${\bf B}({\bf r}_{GVO})$, however the choice of model in this pre-whitening step is of little importance. 
The \textsc{gvo} time series are typically produced at 1-month or 4-month cadence and provided on a global grid of 300 equally spaced locations. 
Comparisons of the \textsc{gvo} time series with independent ground observatory records and spherical harmonic field models (see Figure \ref{Fig: SV GVO comparison}) have demonstrated good agreement on all three vector field components \citep{Hammer_etal_2020a}. The figure shows SV time series, computed as annual differences, at the three ground observatory stations Ascension Island (East component), Honolulu (Radial component) and Hermanus (South component). Note here that the \textsc{gvo} estimates have been mapped to ground in order to ease the visual comparison. \\

\begin{figure*}[!htbp]
\centerline{\includegraphics[width=\textwidth, keepaspectratio=true]{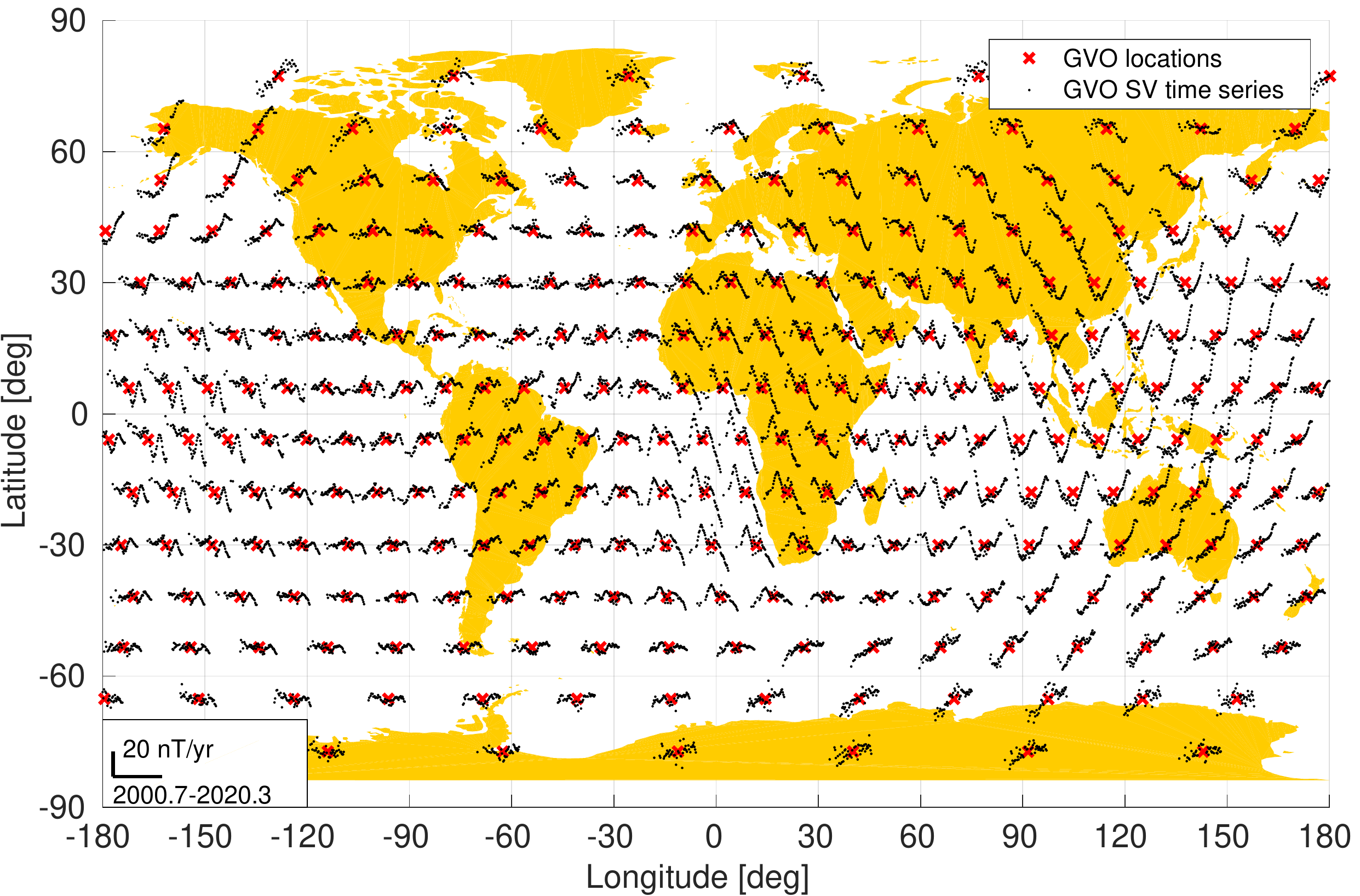}}
\caption{Composite \textsc{gvo} time series of the radial SV ($\partial B_r/\partial t$ -- black dots) over the period 1999--2020, computed as annual differences of 4 monthly \textsc{gvo}s mapped at an altitude of 700km. \textsc{gvo} locations are marked with a red cross. 
The \textsc{gvo} time series have been derived using measurements from Oersted data (1999-2004), \textsc{champ} data (2002-2010), \textsc{chaos}-6x9 calibrated CryoSat-2 data \citep[2010-2014, see][]{Olsen_etal_2020} and data from the {Swarm} trio mission (2014-2020). 
The scale of the time series is shown in the bottom left corner, with the y-axis being 20\,nT.yr$^{-1}$ and the x-axis going from 1999 to 2020. Adapted from \citep{Hammer_etal_2020b}.} 
\label{Fig:GVO_map}
\end{figure*}

\noindent Figure~(\ref{Fig:GVO_map}) presents a map of radial SV 
time series over 1999--2020, at 300 globally distributed \textsc{gvo}s, with 4 month cadence. 
In order to produce such composite time series, the \textsc{gvo}s have been mapped to a common altitude of 700 km. 
This is performed by subtracting from each \textsc{gvo} series the difference between the field at 700 km and at the mean orbital altitude (separately for each satellite mission), as determined from the \textsc{chaos}-7.2 model \citep{Finlay:2020}. 
The map in Figure~(\ref{Fig:GVO_map}) allows the visual inspection of patterns of sub-decadal SV trends and changes. 
Low latitude regions display particularly strong variations, which do not occur simultaneously over all longitudes. 
Distinctive "$\Lambda$" and "$V$" shaped behaviour occur regionally over time spans ranging from 5 to 10 years. 
They can be observed for instance over the South-Atlantic region around 2007 and again in 2014, over Indonesia around 2014, and in the Pacific region around 2017. 
\textsc{gvo}s are thus suitable for studying inter-annual and longer field changes at satellite altitude. 
These likely originate from changes in the outer core fluid motions (see section \S\ref{sec: core physics}). 
The \textsc{gvo} dataset, together with their error estimates, are suitable for core flow inversions \citep{Whaler_Beggan_2015,Kloss_Finlay_2019,Rogers_etal_2019} and data assimilation studies that combine information from geodynamo numerical models with signals observed in the \textsc{gvo} time series \citep{Barrois_etal_2018,huder2019}. 

\section{Geomagnetic field modelling}
\label{sec: modeling}

In order to study core field changes occurring over periods from years to decades, the course of action is typically either to analyse magnetic changes as recorded at a particular location, or to analyse spherical harmonic field models \citep{Mandea_etal_2010}. 
Although the temporal resolution of ground stations is remarkable, their uneven spatial distribution renders global studies of core field changes challenging \citep{Mandea_Olsen_2006}. 
Alternatively, the construction of field models \citep[e.g.,][]{Finlay:2020,Ropp:2020,Sabaka:2020} requires the addition of prior information in order to reduce the non-uniqueness of the geomagnetic inverse problem. 
Models using a spherical harmonic representation come with attractive mathematical benefits -- e.g., orthogonality relations and the addition theorem \citep{olsen2010separation} and have simple means of specifying the model complexity as a function of harmonic degree. 
However, since spherical harmonics are global support functions, it is preferable to estimate such models with an homogeneous data coverage over the Earth. With real data, specific sources dominate the signals over some regions and it is therefore appealing to consider local methods for studying the magnetic field.
We detail in this section local and global methods for modelling the core field.    

\subsection{Local modelling techniques}
\label{sec:SOLA}

Modelling the geomagnetic field is typically done using spherical harmonic expansions, which remains the preferred tool because of their convenient mathematical properties. However, other modelling strategies focusing on local techniques have been proposed, which have typically been developed for modelling the field on regional scales. Local modelling strategies include (i) considering a limited area with a flat-Earth approximation \citep[Rectangular and Cylindrical Harmonic Analysis, see][]{Alldredge_1981,Alldredge_1982}, or with basis functions defined on a spherical cap \citep[Spherical Cap Harmonic Analysis, see][]{Haines_1985} or \citep[Revised Spherical Cap Harmonic Analysis, see][]{Thebault_etal_2004}, (ii) using basis functions such as wavelets \citep{Chambodut_etal_2005, Holschneider_etal_2003}, "quasi-local" functions \citep{Lesur_2006}, Harmonic Splines \citep{Shure_etal_1982, Geese_etal_2010} or Slepian functions \citep{Beggan_etal_2013, Plattner_Simons_2015}, (iii)  using spherical harmonic models with localised constraints \citep{Constable_etal_1993}, (iv) using equivalent source representations such as dipoles \citep{Mayhew_Estes_1983} and monopoles \citep{Hodder_1982, oBrien_Parker_1994}. \\

\noindent Here, we revisit a local approach called Subtractive Optimally Localised Averages (\textsc{sola}). The \textsc{sola} technique allows for localised estimates of the radial field and SV field at the core-mantle boundary (CMB), to be computed as local spatial averages centred on a location of interest \citep{Hammer_Finlay_2019}. These local spatial field averages are computed from averaging kernels which are influenced by the used data. The \textsc{sola} technique provides information on the resolution of the field estimate, in the form of the averaging kernels and variance estimates of the locally averaged field. This information is important when trying to resolve smaller length and time scales of the core field signal. \\

\noindent The approach adopted in geomagnetism builds on the method originally developed in helioseismology by \cite{Pijpers_Thomson_1994}, and is a modified Backus-Gilbert inversion type \citep{Backus_Gilbert_1968,Backus_Gilbert_1970,Pujol_2013}. At satellite altitude, the magnetic vector field can be linked to the radial field at the CMB using the Green's functions of the Neumann boundary value problem \citep{Gubbins_Roberts_1983}. Focusing on the core field, a dark quiet time selection scheme is typically implemented and field model estimates of the lithospheric field, the magnetospheric and ionospheric fields along with their associated Earth induced fields are subtracted \citep{Hammer_etal_2020b}. Since the CMB radial magnetic field is linearly related to the satellite vector data, a \textsc{sola} estimate at a specified target location and time, $(\mathbf{r}_0,t_0)$, at the CMB, can be written as a weighted linear combination of the satellite magnetic data \citep{Backus_Gilbert_1970}:
\begin{equation}
\widehat{B}_r(\mathbf{r}_0,t_0)  = \sum_n^N q_n(\mathbf{r}_0,t_0) \, d_n(\mathbf{r}_n,t_n), \label{eq:SOLA_est}
\end{equation}
where $q_n$ are \textsc{sola} weight coefficients and $d_n$ are the satellite data $(n=1,...,N)$. They are related to the radial magnetic field, $B_r(\mathbf{r}',t_n)$, integrated over the CMB as \citep{Gubbins_Roberts_1983}
\begin{linenomath*}
\begin{equation}
d_{n}(\mathbf{r}_n,t_n) = \oint_{S'} G(\mathbf{r}_n, \mathbf{r}')B_r(\mathbf{r}',t_n) dS' , \label{eq:SOLA_data_int1}
\end{equation}
\end{linenomath*}
where the CMB surface element at radius $r'$ is $dS'=\mathrm{sin}\theta'd\theta'd\phi'$ and the data kernel, $G(\mathbf{r}_n, \mathbf{r}')$, is the spatial derivative of the Green's functions for the exterior Neumann boundary value problem \citep[e.g.,][]{Gubbins_Roberts_1983, Barton_1989}. 
Combining equations \eqref{eq:SOLA_data_int1} and \eqref{eq:SOLA_est}, it is possible to write the \textsc{sola} estimate as
\begin{equation}
\widehat{B}_r(\mathbf{r}_0,t_0)  =  \oint_{S'} \mathcal{K}(\mathbf{r}_0, t_0, \mathbf{r}') \, B_r(\mathbf{r}',t_n)dS'  \text,\label{eq:SOLA_est2}
\end{equation}
where $\mathcal{K}(\mathbf{r}_0, t_0, \mathbf{r}')$ is a spatial averaging kernel. Thus, the \textsc{sola} weight coefficients together with the data kernel, specifies the averaging kernel centered on position $\mathbf{r}_0$ as
\begin{equation}
\mathcal{K}(\mathbf {r}_0, t_0, \mathbf{r}')  = \sum_n^N q_n(\mathbf{r}_0,t_0) \, G(\mathbf{r}_n,\mathbf{r}') \text.
\label{eq:SOLA_kernel_1}  
\end{equation}
To determine the \textsc{sola} weight coefficients, an objective function that describes the departure of the spatial averaging kernel towards a pre-specified target kernel is minimised and information from a data error covariance matrix is included \citep{Hammer_Finlay_2019}. By varying the target kernel width, \textsc{sola} estimates of different spatial averaging width (i.e. resolution) emerges. Typically, a Fisher distribution is used as a target kernel \citep{Fisher_1953}, however other options are possible \citep{Masters_Gubbins_2003}.\\

\noindent When calculating \textsc{sola} field estimates, data from within a month are commonly used and time-dependence neglected. In order to compute \textsc{sola} SV estimates, data from within a window (typically 2 years) are used, where the time-dependence is handled using a first order Taylor expansion assumed valid close to a reference time \citep{Hammer_etal_2020b}. By taking differences between consecutive \textsc{sola} SV field estimates, estimates of the SA at the CMB can be derived, which is of interest when studying core dynamics \cite[e.g.,][]{Finlay_etal_2016a, chiduran2020}. In particular, using the high quality measurements from the {Swarm} mission, it is possible to take 1-yr differences of \textsc{sola} SV estimates derived from 1 year data windows \citep{Hammer_etal_2020b}.\\

\begin{figure*}[htbp]
\centerline{
\subfloat[\textsc{sola} radial SA, 1yr time windows, 1 month steps.]{\includegraphics[angle=0, width=7.cm]{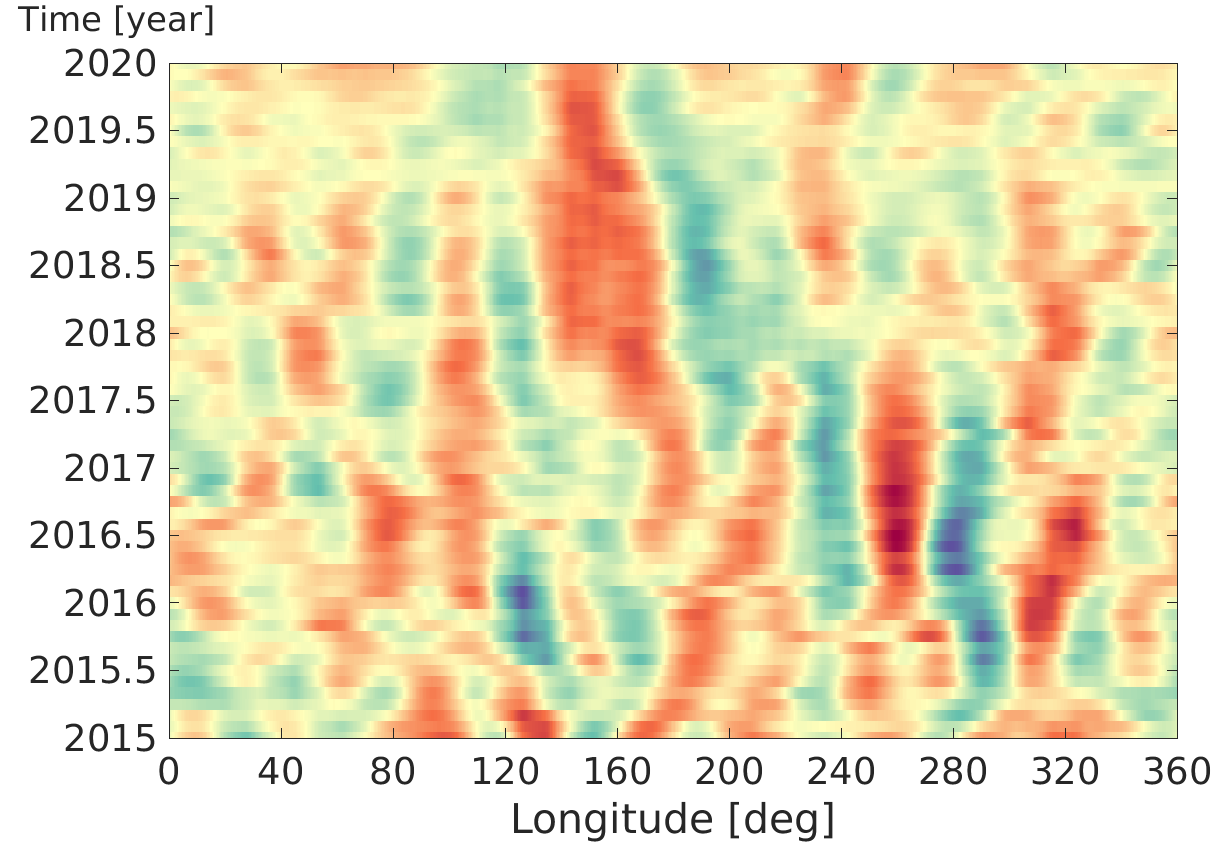}}\hspace{0.1cm} 
\subfloat[\textsc{chaos}-7.2 radial SA.]{\includegraphics[angle=0, width=7.cm]{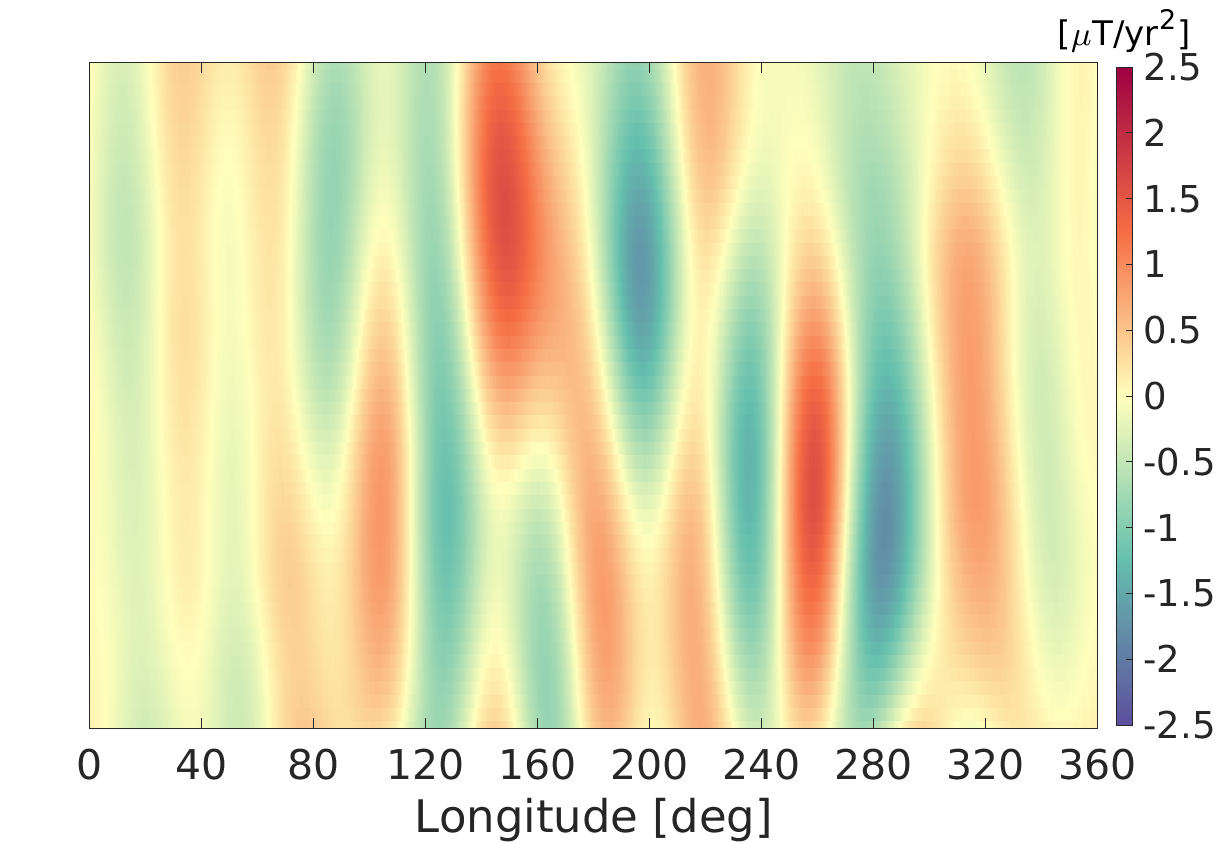}}}
\caption{Time-longitude plots of the radial SA field along the CMB geographical equator, showing: 1yr differences of \textsc{sola} SV estimates derived from 1-yr data windows moving in 1 month steps (left), and predictions from the \textsc{chaos}-7.2 model for spherical harmonic degrees 1 to 10 (right). Adapted from \citep{Hammer_etal_2020b}.}
\label{Fig:TL1yr}
\end{figure*}

\noindent Because the \textsc{sola} method allows for local field estimates to be determined, collections of individual \textsc{sola} estimates having the same resolution, can be combined into making regional or global maps.
Of particular interest for more detailed studies, are regions of strong fluctuations in the field changes observed at low latitudes. In order to examine such changes, the \textsc{sola} method can be used for making SA estimates along the CMB equator in consecutive time steps, such that the evolution of the SA can be mapped. Figure (\ref{Fig:TL1yr}) presents time-longitude (TL) plots of the SA evolution on the geographic equator (centred on the Pacific region) at the CMB between 2015.0 and 2020.0 derived from the radial field measurements of the {Swarm} satellites. The left plot shows \textsc{sola} SA estimates derived from 1-yr differences of \textsc{sola} SV determined from 1 year data windows, sliding in 1 month steps, while the right plot shows the SA predictions from the \textsc{chaos}-7.2 model for spherical harmonic degrees 1 to 10 -- i.e. a maximum wavelength $\sim$2200km at CMB. Here the \textsc{sola} SA estimates have associated errors estimates of about $0.6$ nT.yr$^{-2}$ and averaging kernel widths of $42^{\circ}$, corresponding to length scales of 2500km at the CMB. Comparing the two TL-plots, similar large scale features of the SA evolution with amplitudes of $\pm 2.5$ nT.yr$^{-2}$, can be observed in both the \textsc{sola} and \textsc{chaos} plots. However, the \textsc{sola} plot demonstrates increased temporal resolution, and reveals coherent patterns that have been smoothed out by the temporal regularisation of the \textsc{chaos}-7.2 model. \\

\noindent In the \textsc{sola} TL plot, interesting structures includes intense SA patches under Central America, at longitudes from $240^{\circ}$ to $320^{\circ}$, observed going from 2015.5 to 2018. Also observed are strong side-by-side positive and negative SA features emerging in the Pacific region around 2017, at longitudes $150^{\circ}$ to $220^{\circ}$, which are seen to have drifted approximately $20^{\circ}$ westwards until 2020. Such rapid variation at low latitudes may be caused by time variations of the liquid outer core flow, and could provide important constraints on the equatorial dynamics of the outer core \citep{Kloss_Finlay_2019}. Responsible mechanisms for such flows may involve equatorially-trapped MAC waves in a stratified layer at the core surface \citep{Buffett_Matsui_2019, chiduran2020}, Magneto-Coriolis modes \citep{gerick2020} or equatorial focusing of hydrodynamic waves related to turbulent convection deep within the core \citep{aubert2019geomagnetic}. Possible mechanisms are described in more details in section~\S\ref{sec: core physics}.

\subsection{Global field models}
\label{sec:gfm}

Although building localised representations of the core magnetic field presents numerous advantages, it still relies on a rough separation of the different sources often based on outputs of other's modelling efforts. 
We therefore present here global magnetic field modelling techniques that simultaneously consider the core and other contributions. 
Although different types of representation are possible, magnetic field components are most generally described as the negative gradient of potentials of internal or external origins, respectively $V_i$ and $V_e$, 
\begin{equation}
{\mathbf B}(\theta,\phi,r,t)= - \nabla \{ V_i(\theta,\phi,r,t) + V_e(\theta,\phi,r,t) \} \,,
\end{equation}
themselves being decomposed into spherical harmonic as
\begin{equation}
\begin{array}{ll}
\displaystyle V_i(\theta,\phi,r,t) = a \sum_{\ell=1}^{L_i}  \sum_{m=0}^{m=\ell}\; \{ g_\ell^m(t)  \, \cos(m \phi) +  h_\ell^m(t)  \, \sin(m \phi) \}\; \left(\frac{a}{r}\right)^{\ell+1} & \displaystyle P_\ell^m(\cos \theta)\,,\\
\displaystyle V_e(\theta,\phi,r,t)=  a \sum_{\ell=1}^{L_e}  \sum_{m=0}^{m=\ell}\; \{ q_\ell^m(t)  \, \cos(m \phi) +  s_\ell^m(t)  \, \sin(m \phi) \}\; \left(\frac{r}{a}\right)^{\ell} & \displaystyle P_\ell^m(\cos \theta)   \text.
\end{array}
\label{eq:pfields}
\end{equation}
Here, $(\theta,\phi,r)$ are colatitude, longitude and radius in geocentric system of coordinates. The $P_\ell^m(.)$ are Legendre functions. The $g_\ell^m$, $h_\ell^m$, $q_\ell^m$ and $s_\ell^m$ are the Gauss coefficients to be estimated independently for each source, where $h_\ell^0$ and $s_\ell^0$ are null. 
All these Gauss coefficients depend on time $t$. 
In such a decomposition it is assumed that there are no sources of magnetic field at the measurement point, which usually is true for ground measurements, but it is not necessarily valid at satellite altitude, in particular in the auroral regions with field-aligned currents \citep[see][]{finlay2017challenges}. 
The internal and external description refers to a sphere of radius $a=6371.2$km, that is an approximation of the Earth's surface. 
Note that with satellite data, ionospheric fields are generated in between the measurement position and the Earth's surface, introducing some confusion in this simple description.
The same difficulty in principle exists for the lithospheric field at the equator, that is clearly outside the reference sphere, but this latter effect is generally neglected.\\

\begin{figure}
{\center
\includegraphics[width=7.cm, angle=0,keepaspectratio=true]{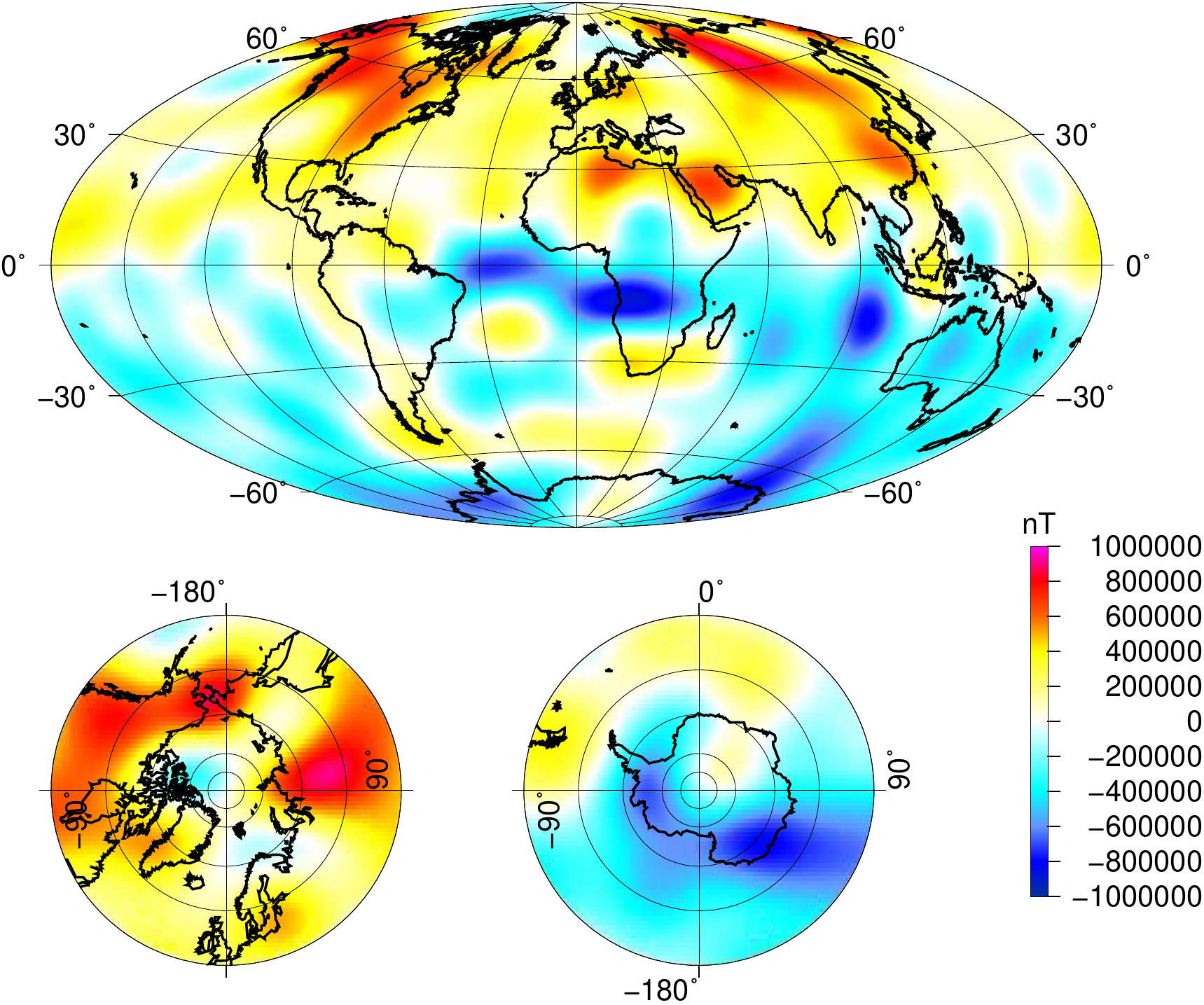}
\includegraphics[width=6.8cm, angle=0,keepaspectratio=true]{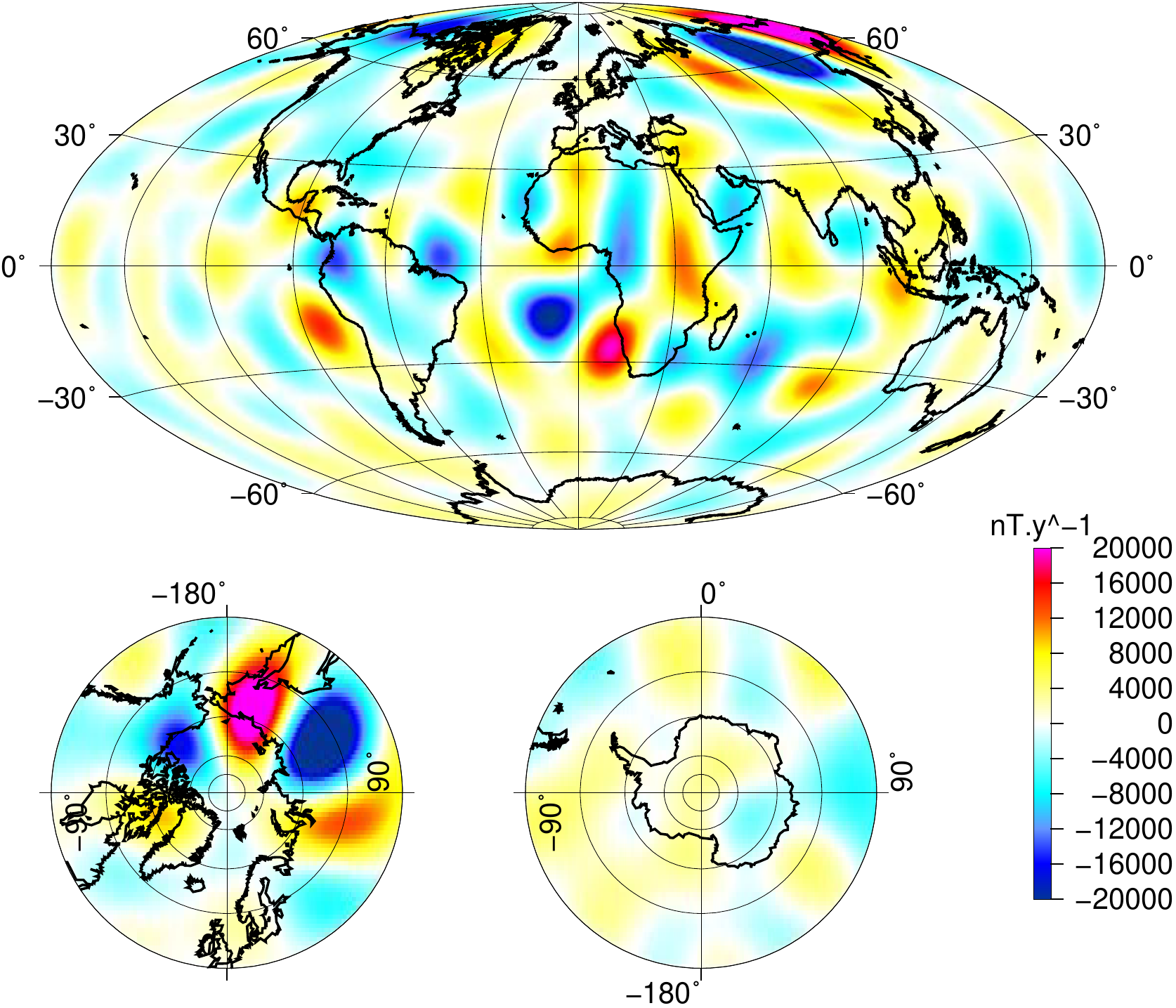} 
\caption{Vertical down core magnetic field as estimated by the \textsc{mcm} model \citep{Ropp:2020} truncated at spherical harmonic 13 (left) and its secular variation (right). The field has been mapped at the CMB for epoch 2015.0. In both cases continents are drawn to help location.}
\label{fig:csfield}
}
\end{figure}

\noindent Over the 20 last years, the nearly continuous series of magnetic satellite missions allow to build particularly high resolution, time dependent models of the magnetic field. 
Examples of these models are the Comprehensive models \citep{Sabaka:2016, Sabaka:2020} 
, \textsc{chaos} models  \citep{Olsen:2006, Finlay:2020}, GRIMM models \citep{Lesur:2008, Lesur:2014}, for the most known. 
This list is far from being exhaustive. 
Special issues associated with the publication of the IGRF  generally give a better picture of the diversity and numbers of magnetic models available \citep[see e.g.,][]{Maus:2005d, Finlay:2010b, Thebault:2015b, alken2020evaluation}. 
If these models are different, they agree overall on the core magnetic field and its SV strength and direction at the Earth's surface -- see \cite{alken2020evaluation} for more information on statistics of the differences. 
An example is displayed in Figure~\ref{fig:csfield}, for the epoch 2015.0, of the vertical down components at the CMB from the \textsc{mcm} model  \citep{Ropp:2020}.
It should be noted that in these images, the spherical harmonic degree is limited to $\ell=$13, and that shorter wavelength signals, that can be dominent and therefore strongly modify the images, are not included. \\

\begin{figure}
{\center
\includegraphics[width=7.cm, angle=0,keepaspectratio=true]{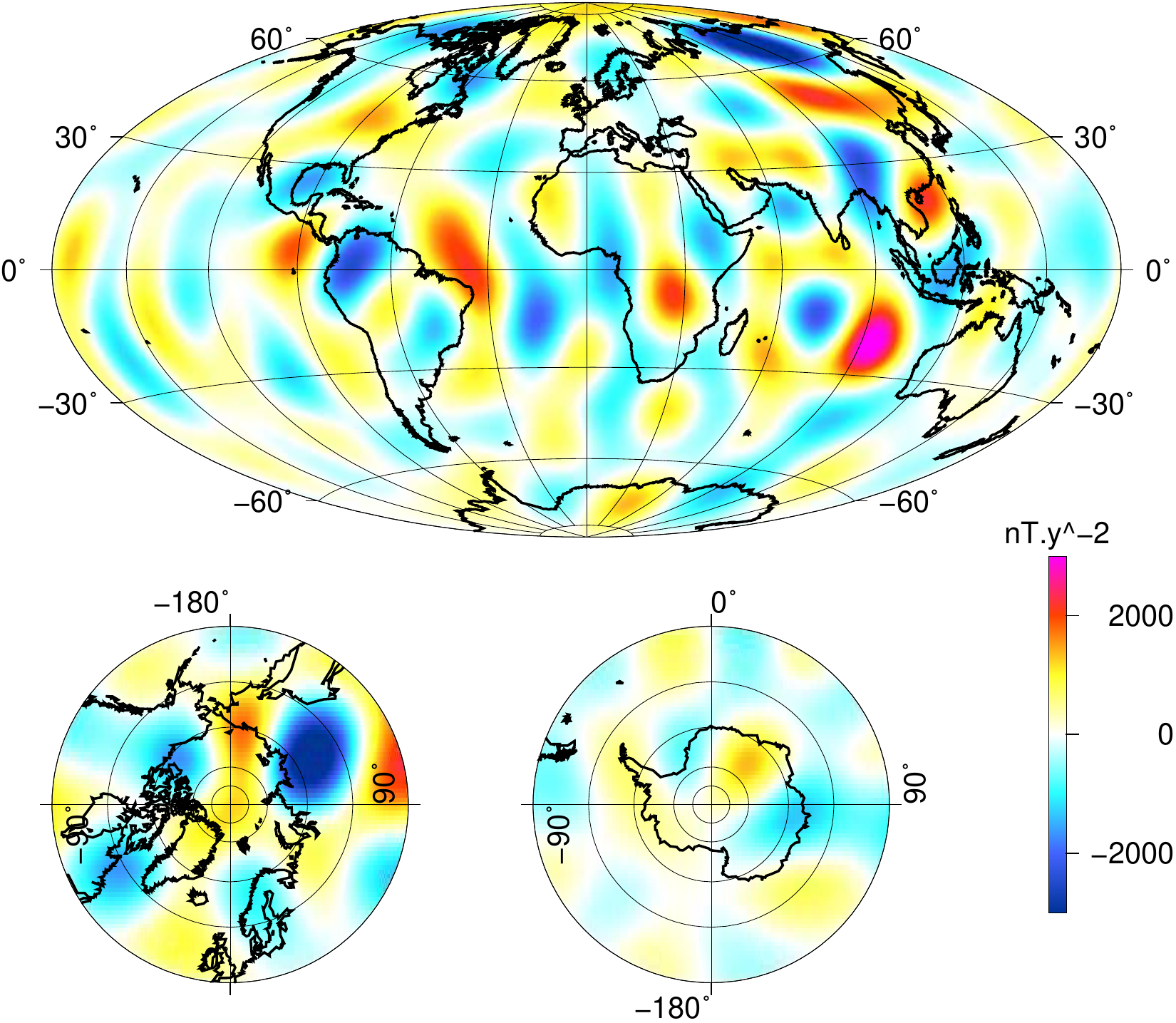}
\includegraphics[width=7.cm, angle=0,keepaspectratio=true]{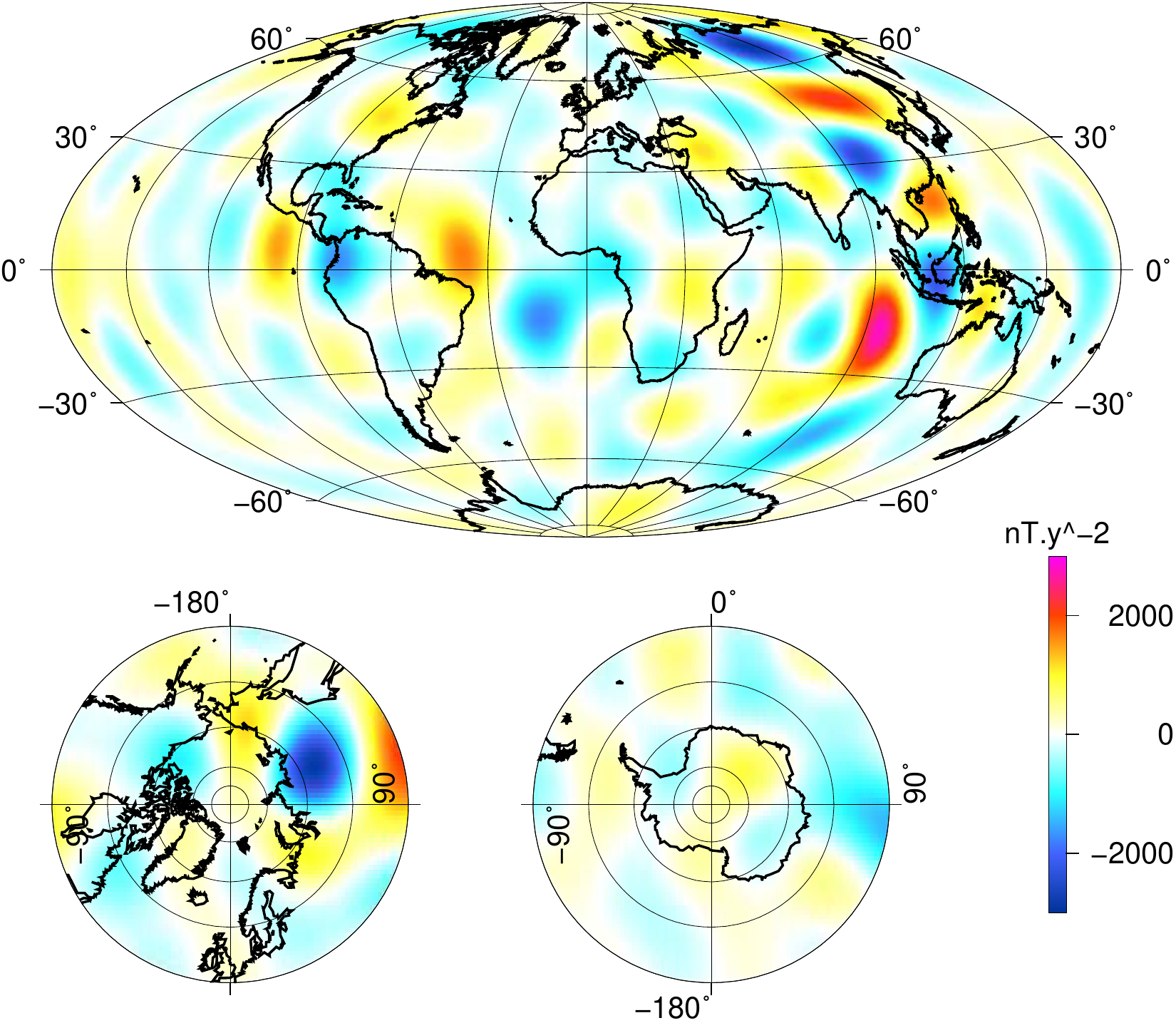} \vspace{0.5cm}\\
\includegraphics[width=7.cm, angle=0,keepaspectratio=true]{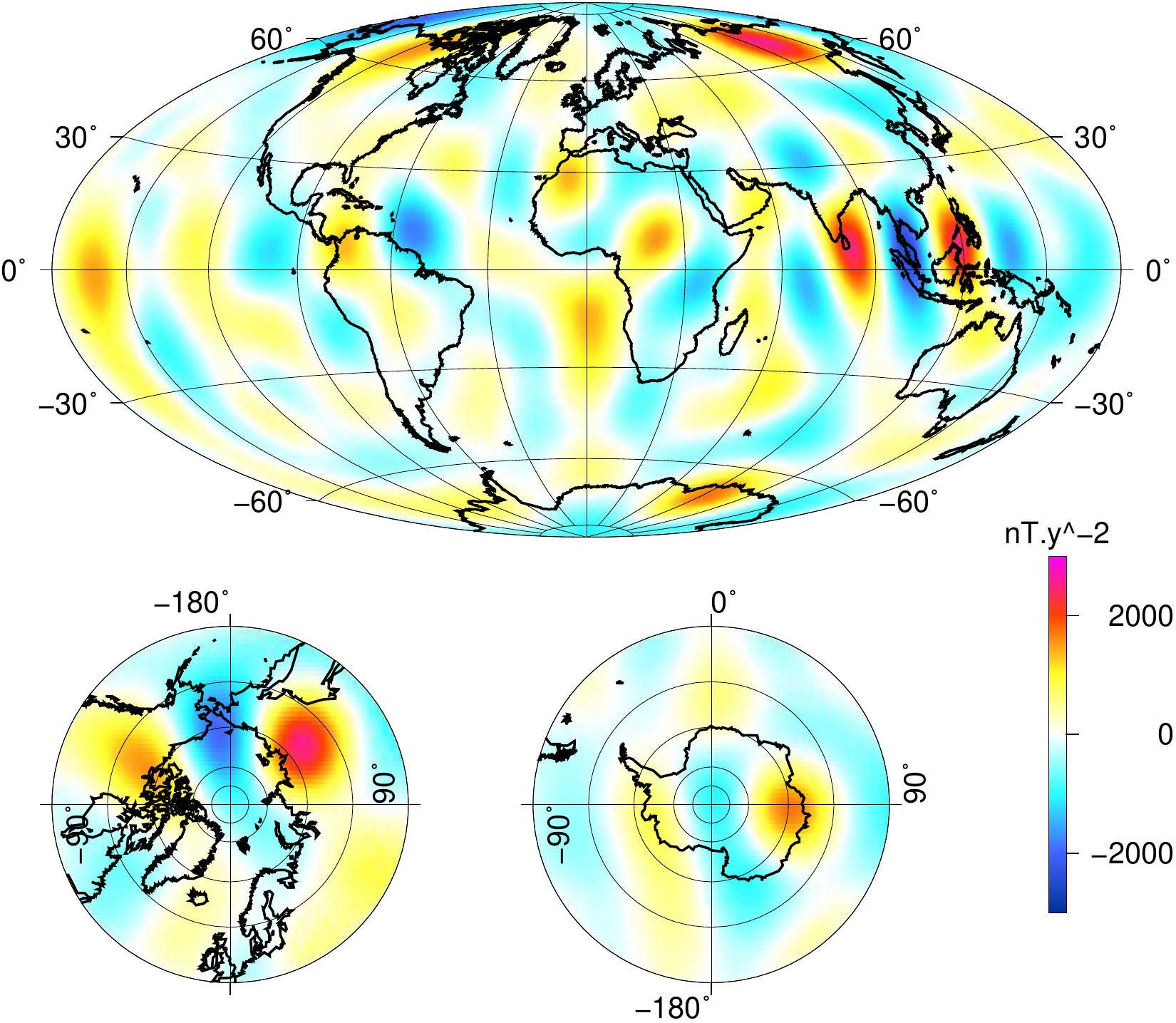}
\includegraphics[width=7.cm, angle=0,keepaspectratio=true]{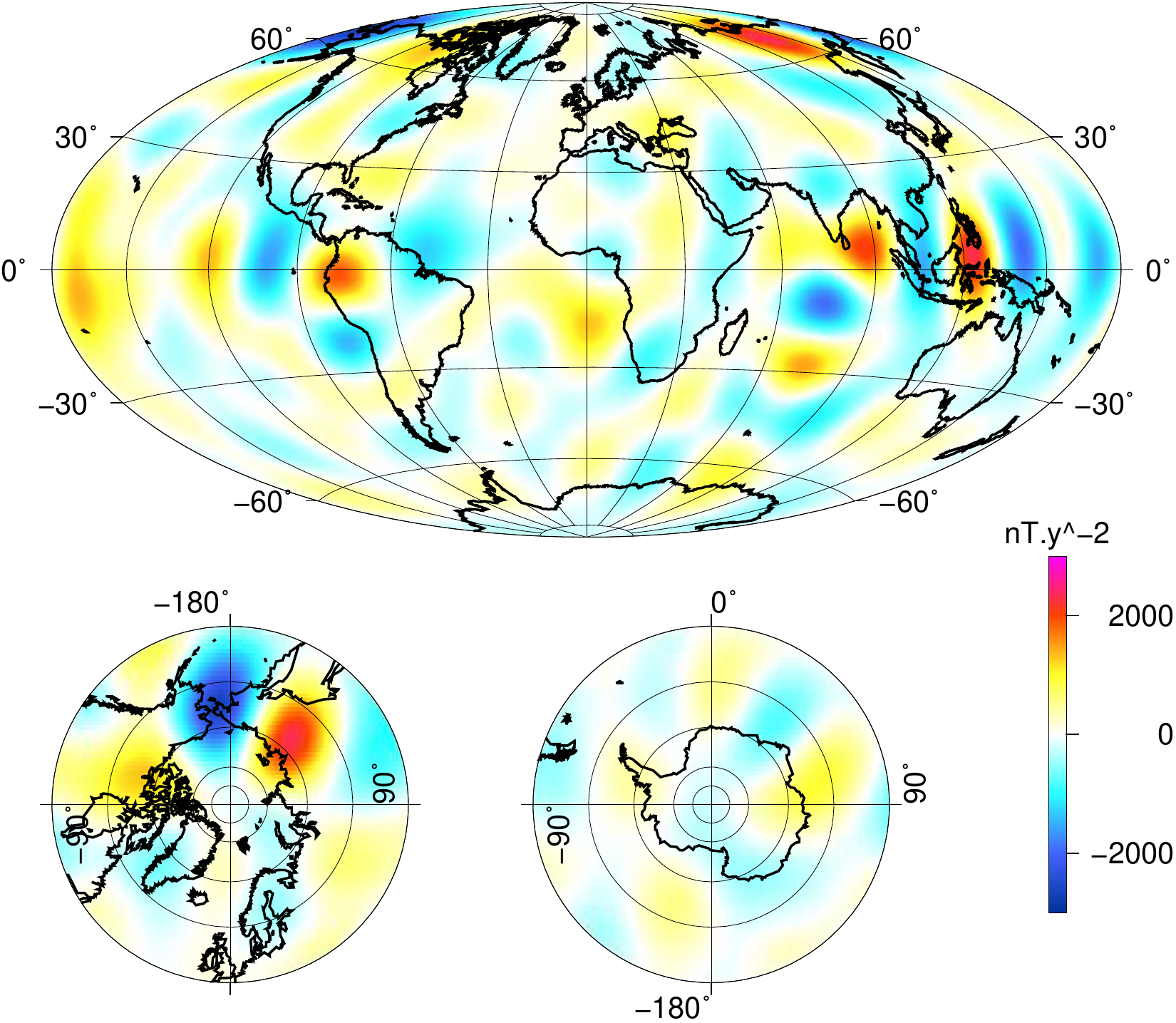}\\
\caption{Vertical down core magnetic field acceleration as estimated by the \textsc{mcm} model \citep{Ropp:2020} (left) or \textsc{chaos}-7 model \citep{Finlay:2020}  (right), truncated at the spherical harmonic $\ell=13$. The field has been mapped at the CMB for year 2006 (first row) and 2018 (second row). Continents are drawn to help location.}
\label{fig:afield}
}
\end{figure}

\noindent The agreement between models for the SA remains generally good but they present significant differences in terms of amplitudes and locations of the acceleration patterns. Maps of the vertical down SA component for 2006 and 2018, at the CMB, are displayed in Figures~(\ref{fig:afield}) for two models \textsc{chaos}-7 \citep{Finlay:2020} and \textsc{mcm} \citep{Ropp:2020}. As for the SV, the spherical harmonic degree is limited to $\ell=$13, and shorter wavelength signals can significantly modify these maps. Differences between the two models are obvious even if there is a general agreement that strong acceleration patterns are mainly located at the CMB along the equatorial regions, or at all latitudes in between 90 and 120 degree East. Strongest accelerations are often seen under Indonesia or Central America. These acceleration patterns have a footprint at the Earth's surface that is very large, in agreement with those directly observed in Figure~(\ref{Fig:GVO_map}). 

\noindent 

\noindent The observed differences between models result from the way the separation of sources contributing to the measured magnetic field is achieved.
The coverage in space and time of the combined observatory and satellite data sets is such that for temporal scales of the order of a few days to a few weeks, the separation of large scale internal fields from external fields is suitable for core field modelling. 
There may be difficulties for short spatial wavelengths, or short temporal scales but these are generally irrelevant for the core field. One can refer to \cite{Lesur:2012} for a description on how fast field variations in time leaks in small wavelength of a nearly static internal field. Satellite data are also contaminated by non-potential magnetic fields (i.e. magnetic field that cannot be described as gradients of potentials as in equations~(\ref{eq:pfields})). These signals are mainly due to field aligned electric currents linking the magnetosphere to the ionosphere. These non-potential fields can be co-estimated with other sources \citep[e.g.,][]{Sabaka:2016, Lesur:2008}, although, even if not modelled, it seems that their contributions do not significantly affect the core field models derived from satellite data.
Therefore, the main difficulties that remain for precise core field models are the separation of sources that are seen as internal from the satellite altitude. \\

\noindent From the satellite altitude down, there are firstly the magnetic fields generated in the ionosphere during night-times
(day-time data are not usually used for core field modelling). There have been several attempts to model these fields, particularly at high latitudes \citep[e.g.,][]{Lesur:2008}, but they remain difficult to handle because satellites generally fly at nearly sun-synchronous orbits and therefore over a few days acquire data only on a narrow local time window. It should be noted that because the Earth rotates under the ionosphere that is locked in local time, the ionospheric signals may preferentially affect zonal Gauss coefficients (i.e. $g_l^m$ and $h_l^m$ with $m=0$) of the core contribution.
Secondly there are the fields generated in the lithosphere. So far, there has been no successful attempt to separate the lithospheric field from the core field. 
It follows that the contribution of the lithospheric field is ignored at long wavelengths. At shorter wavelengths, from spherical harmonic degree higher than $\ell=$16, where the lithospheric field is dominent at the Earth's surface, the static core field contribution is unknown. 
At the same altitude, signals generated in oceans can be well estimated when linked to tides because of their well established periodicities \citep{Grayver:2019}. However, the signals associated with general oceanic circulation have not been isolated yet. 
There are probably other sources in the crust that have not been identified, such as possibly the water cycle that has a strong signature in gravity data. 
However, the field induced in the conductive crust and mantle by the time varying external fields remain the major difficulty that precludes the derivation of accurate core field models as both contributions overlap over a large range of spatial and temporal scales \citep[see e.g.,][]{olsen2005ionospheric}. It is presently the isolation of long periods external fields (and their induced counter-parts) that sets the temporal resolution of core field models. Ambiguities may remain at the decadal period of the solar cycle (and its harmonics), and more importantly we lack accurate understanding of the strong semi-annual and annual periodicities that arise naturally in ionospheric and external fields.\\

\begin{figure}
{\center
\includegraphics[angle=0, width=10cm, keepaspectratio=true]{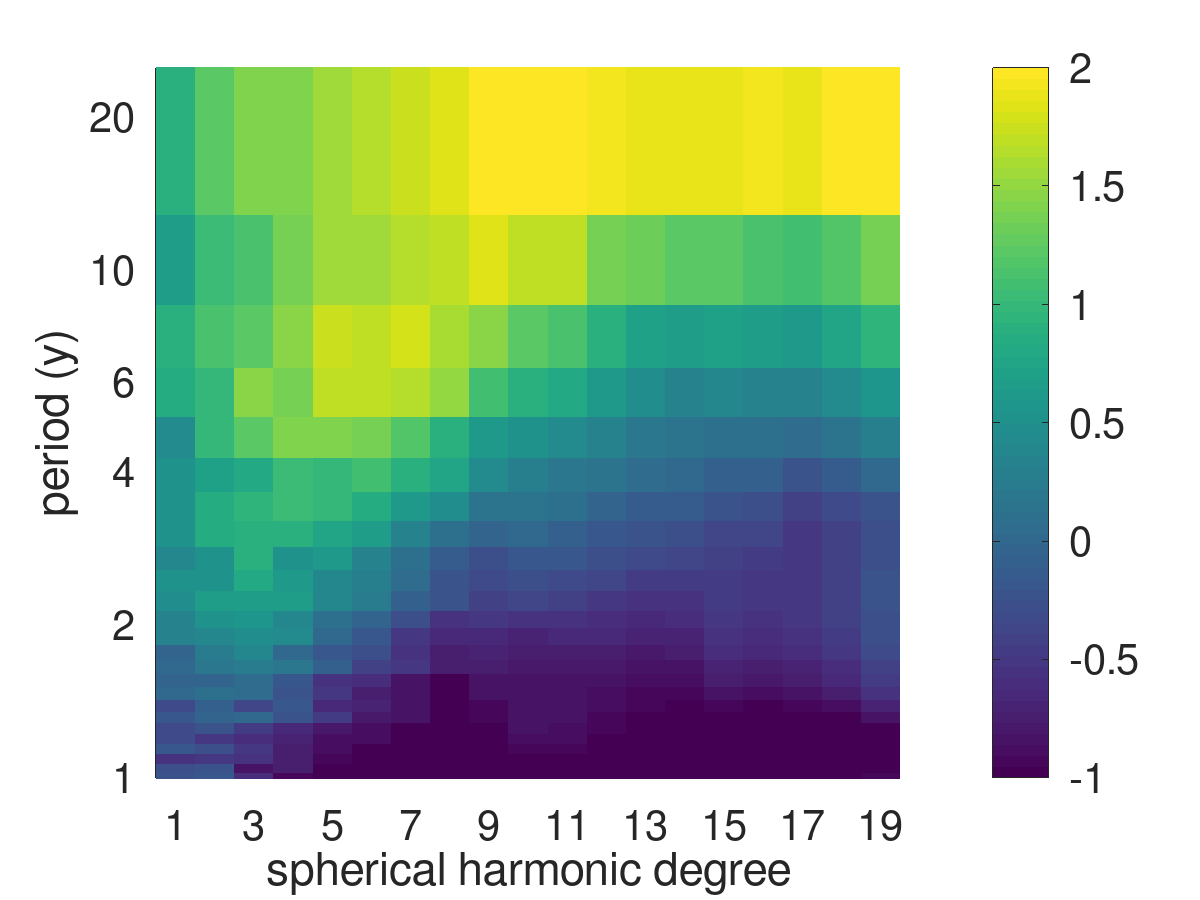}
\caption{Intensity of the SA at the core surface ($\log_{10}$ of nT.yr$^{-2}$) as a function of the period and the spherical harmonic degree, for the \textsc{chaos}-7 field model \citep{Finlay:2020}. Adapted from \cite{gillet2019}.
} 
\label{Fig:SA chart}}
\end{figure}

\noindent Generally the strategy used to separate these various "internal" fields from the core field is based on a careful data selection, a partial modelling of the different contributions, and often, a constraint (i.e. a damping or regularisation) set on the core field contribution such that it does not vary too rapidly in time. 
This approach has been used for a long time, and examples of application to satellite data can be found in e.g.,  \cite{Olsen:2006, Lesur:2008} or more recently \cite{Sabaka:2020, Finlay:2020}. 
The characteristic of this approach is that the core magnetic field SV and SA are averaged over time. Furthermore, their temporal resolutions vary depending on the spatial scale. 
As shown in Figure~(\ref{Fig:SA chart}), with these approaches the acceleration is likely underestimated at short periods and small wavelengths -- i.e. from spherical harmonic degree $\ell=5$ up.
An alternative approach has emerged, pioneered by \cite{McLeod:1996, Gillet:2013}, where the temporal constraints applied to construct the core field model is derived from what is known of the spatial and temporal behaviour of each source contributing to the observed magnetic field. 
The temporal constraints are imposed either through a priori time cross-covariances \citep{huder2020cov}, or by time-stepping associated stochastic equations \citep{huder2019, Baerenzung:2020, Ropp:2020}.  
As a result, the modelled SV and SA are not averaged over time by construction, to the cost of having models with large posterior uncertainties \citep{Gillet:2013,Holschneider:2016}.  We point for instance to \cite{Ropp:2020} where outputs from numerical dynamos are used as prior spatial information on the core field behaviour.
Whatever modelling technique is used, some temporal averaging may still exist in the models due to the imperfect data coverage. In particular the nearly continuous satellite era extends only over the last 20 years precluding robust estimation of temporal periodicities in the core exceeding 10 years.\\

\noindent All these modelling techniques rely strongly on what is assumed by the modeller regarding the magnetic field generated by different sources, and in particular by the Earth's core. 
To better understand the dynamics of the liquid outer core, it is therefore important to define how fast the core field varies with time at the Earth's surface. 
The fastest variations in the core field are necessarily smoothed out by (i) magnetic diffusion within the fluid core and (ii) induction in the conductive mantle. 
However, the associated cut-off periods are not well known. 
On the core side, the most extreme numerical simulations of the geodynamo suggest that it could be of the order of 1 year. This is hard to confront immediately to magnetic data because of the dominant external signals towards high frequencies (see Figure~(\ref{fig:FFT})). 
On the mantle side, the cut-off period remains difficult to estimate because of the poorly constrained lowermost mantle conductivity \citep[e.g.,][]{Kuvshinov:2021}. This cut-off value is likely less than 1 year \citep[see][]{jault2015conductivity}. 
It is interesting to see how much variability has been accepted for the core field or its SV in the currently available models. 
Both quantities are characterised by the SV and SA time scales, as defined by respectively
\begin{equation}
\tau_{sv}(\ell) = \sqrt{\frac{\sum_{m} (g_\ell^m)^2 + (h_\ell^m)^2 }{\sum_{m} (\dot g_\ell^m)^2 + (\dot h_\ell^m)^2}}
\label{eq:tausv}
\end{equation}
and
\begin{equation}
\tau_{sa}(\ell) = \sqrt{\frac{\sum_{m} (\dot g_\ell^m)^2 + (\dot h_\ell^m)^2 }{\sum_{m} (\ddot g_\ell^m)^2 + (\ddot h_\ell^m)^2}}\,.
\label{eq:tausa}
\end{equation}
In these equations we use the notations $\dot{x}=\partial x/\partial t$ and $\ddot{x}=\partial^2 x/\partial t^2$. 
From the combined analysis of magnetic field models and numerical simulations, \cite{Christensen:2012} propose,  for $1 < \ell \le 13$, $\tau_{sv}(\ell) \simeq 480/\ell$ \citep[in fair agreement with][]{lhuillier11} and $\tau_{sa}(\ell) \simeq 11$ years for $\ell\le10$.
With the new generation of numerical geodynamo simulations \citep{Aubert:2018}, the $\tau_{sa}$ value decreases down to sub-decadal periods (for the largest length-scales) as conditions closer to Earth-like are reached \citep{aubert2021}. \\

\begin{figure}
{\center
\includegraphics[width=10.cm, angle=270,keepaspectratio=true]{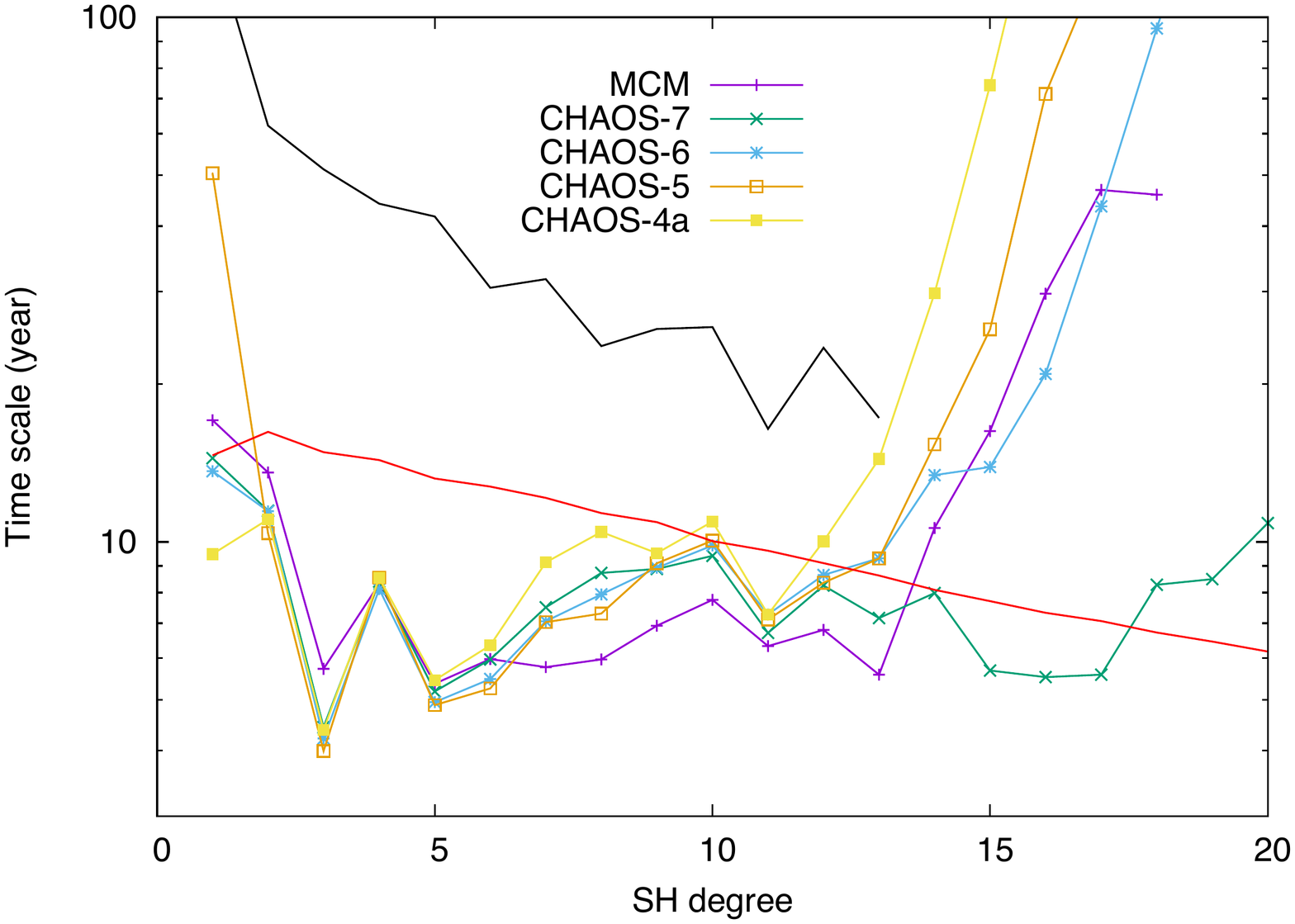}
\caption{Acceleration time scales $\tau_{sa}(\ell)$ as a function of spherical harmonic degree $\ell$, for year 2006, and as estimated with successive versions of \textsc{chaos} field model \citep{Finlay:2020} and the \textsc{mcm} model \citep{Ropp:2020}. Are also shown a characteristic time scale derived from the Coupled Earth numerical dynamo model \citep[red line --][]{aubert2013bottom}, and a time scale obtained neglecting the flow acceleration in the core and the diffusion (black line -- see the main text).} 
\label{fig:timescale}}
\end{figure}

\noindent Because the SV energy is relatively well constrained by observations, $\tau_{sa}$ depends directly on the SA energy. 
The latter can be decomposed, from the induction equation in the fluid outer core
\begin{equation}
\dot {\mathbf B} = \nabla \times ( {\mathbf u} \times {\mathbf B}) + \eta \Delta {\mathbf B} \,,
\label{eq:de}
\end{equation}
 into three terms as
\begin{equation}
\ddot {\mathbf B} = \nabla \times (\dot {\mathbf u} \times {\mathbf B}) +  \nabla \times ({\mathbf u} \times \dot {\mathbf B}) + \eta \Delta \dot {\mathbf B} \,. 
\label{eq:dde}
\end{equation}
Here ${\mathbf u}$ is the flow in the Earth's liquid outer core, and $\eta$ the magnetic diffusivity. 
The first term in the righthand side of equation~(\ref{eq:de}) is the SV generated by the advection of the magnetic field by the flow, 
whereas the second term is linked to magnetic diffusion. 
In equation~(\ref{eq:dde}), the first term in the righthand side is the SA generated by the flow acceleration, the second is the SA generated by the flow advecting the SV, and the last term is the contribution from diffusion. 
Results from numerical dynamo experiments suggest that the two first terms in equation~(\ref{eq:dde}) strongly dominate the last one, and that the first term most often dominates over the second \citep{Aubert:2018,aubert2021}. 
The second term of the same decomposition, neglecting diffusion, was used by \cite{Lesur:2009} to estimate a lower bound for the acceleration energy. Presented in Figure~(\ref{fig:timescale}), it can be seen as an upper limit for an acceptable acceleration time scale $\tau_{sa}(\ell)$. All recent magnetic field models have acceleration time scales for spherical harmonic degree up to $\ell=13$ well below this limit. \\

\noindent In the same figure is shown $\tau_{sa}(\ell)$ derived from the Coupled-Earth numerical dynamo model \citep{aubert2013bottom}. 
For this dynamo simulation \citep[as for the ones used by][]{Christensen:2012} transient phenomena such as SA pulses are excluded because of a too weak magnetic to kinetic energy ratio. 
This tends to produce larger estimates of $\tau_{sa}$ (see also Section \S\ref{sec: core physics} for possible physical interpretations of SA pulses). 
Therefore, except for spherical harmonic degree $\ell=$1 (where the Coupled-Earth dynamo model is known to present too weak SV), it is expected that the estimates for $\tau_{sa}$ from field models remain below the one derived from the Coupled-Earth model. 
As seen in Figure~(\ref{fig:timescale}), this weak constraint is respected by recent magnetic field models (as shown for the \textsc{chaos}-7 and \textsc{mcm} models). 
For the \textsc{chaos} series of models, the progress in modelling and the precise adjustment of the smoothing constraints led to a significant decrease of the time scales towards large spherical harmonic degrees. 
The \textsc{mcm} model that uses alternative temporal representation (see above), presents even lower time scale for degrees $\ell$ from 6 to 13, but slightly larger values at the lowermost degrees.
 It is reassuring that all models seem to agree for intermediate length-scales ($\ell$ around 5). 
Clearly, above degree $\ell=$13 estimates for $\tau_{sa}$ are poor for all models (the SA power drops as it is badly constrained). 
However, there is no insurance yet that at lower spherical harmonic degrees the time scales estimates are exact. 
It is possible that some fast variations of the core field are accounted for within the induction part of the models, or that rapid induced and ionospheric signals are modelled as generated in the core. 
We note however, that the typical 11-yrs value for the acceleration time scale obtained in \cite{Christensen:2012} seems overestimated, in agreement with the latest simulations of the geodynamo.\\ 
 
\begin{figure}
{\center
\includegraphics[width=7.9cm, angle=0, keepaspectratio=true]{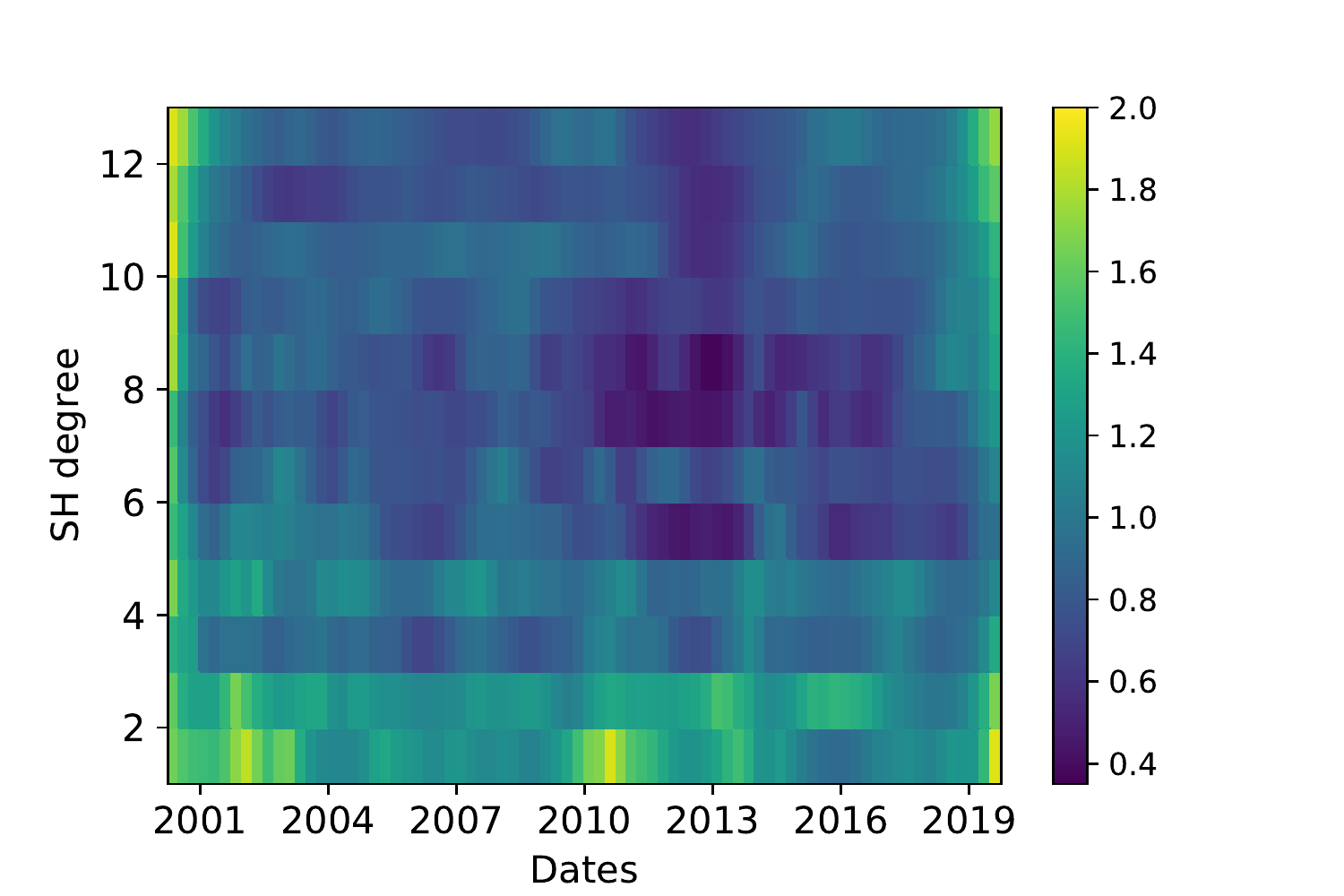}
\includegraphics[width=7.9cm, angle=0, keepaspectratio=true]{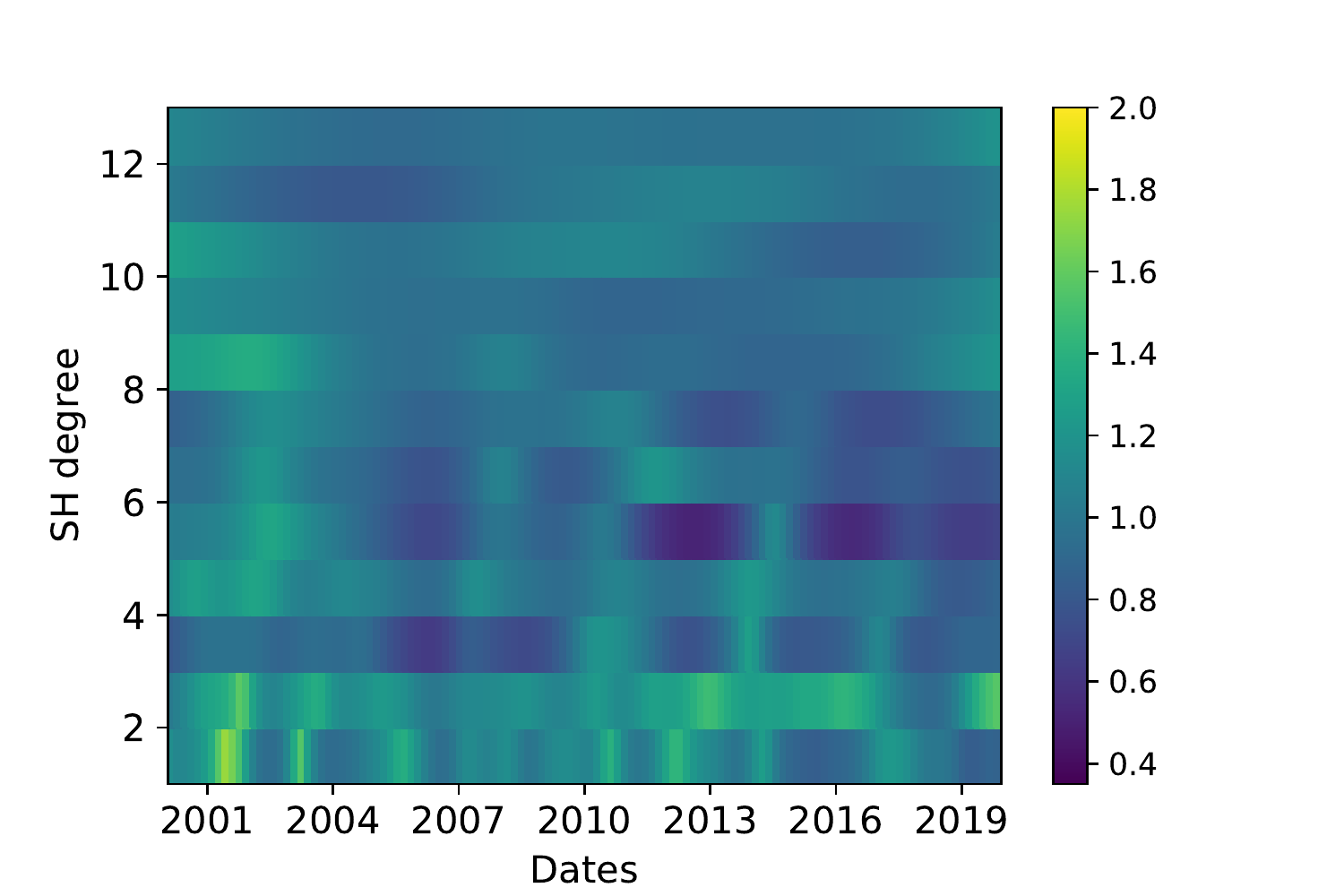}
\caption{Time scales from 2000 to 2020, for a spherical harmonic degree range $\ell \in [1;13]$ (units in $\log_{10}$ of years). Time scales are shown for the \textsc{mcm} model (left) and \textsc{chaos}-7 model (right).} 
\label{fig:timescale-map}}
\end{figure}

\noindent These time scales are varying with time over the satellite era. 
 Figure~(\ref{fig:timescale-map}) presents temporal evolution of $\tau_{sa}(\ell)$ derived from the \textsc{mcm} model \citep{Ropp:2020} and the \textsc{chaos}-7 model \citep{Finlay:2020} from 2000 to 2020. 
Edge effects are visible for the \textsc{mcm} model in 2000 and 2020. 
Both models show a significant variability up to spherical harmonic degree $\ell = 6$, above which the \textsc{chaos}-7 SA appears smoother. 
However, values obtained with \textsc{mcm} are associated with large uncertainties. 
A good agreement in terms of variability is found for degrees 3, 4 and 5. 
Discrepancies are important for the spherical harmonic degree $\ell=1$ where part of the variability coincides with satellite data availability. 
Much more variability is obtained with the \textsc{chaos}-7 model, and this difference with the \textsc{mcm} model is probably due to different way of handling high latitude night-time ionospheric signal, and/or induced fields.\\

\noindent  As described above, the variability of the magnetic field is a key information to better understand the outer core dynamics. 
Progresses have been tremendous in terms of spatial and temporal description of this variability. 
In particular this leads to the possibility to identify waves and other transient phenomena in the core.

\section{Possible sources to inter-annual geomagnetic changes}
\label{sec: core physics}

The apparent periodicity of SA pulses \citep{chulliat2014geomagnetic,Finlay_etal_2016a} has motivated the search for waves as an origin to rapid SV changes. 
In the presence of global rotation the fluid response to transient perturbations consists of inertial waves whereas, within an ambient magnetic field, the responses to perturbations are made of Alfv\'en waves.
For the former waves the restoring force comes from the Coriolis force, while for the latter it is the Lorentz force \citep[for a review, see][]{finlay2010short,Jault2015TOG}. 
Their typical periods are comparable with the rotation and Alfv\'en times, respectively:
\begin{equation}
\tau_{\Omega}=1/2\Omega \mbox{ and } \tau_{A}=c/V_A\,,
\end{equation}
with $\Omega\simeq 7.3\times 10^{-5}$ rad.s$^{-1}$ the Earth's rotation rate, $c=3485$ km the core radius, and $V_A=B/\sqrt{\rho\mu}$ the Alfv\'en speed ($\rho\simeq10^4$ kg.m$^{-3}$ is the core density and $\mu=4\pi\times10^{-7}$ H.m$^{-1}$ the magnetic permeability). 
For a magnetic field intensity $\approx 5$ mT within the core \citep{Gillet:2010N}, one gets $\tau_{A}\approx 2$ years that is much longer than $\tau_{\Omega}$, as measured by the Lehnert number $Le = \tau_{\Omega}/\tau_A\approx 10^{-4}$. 
For planetary scale Alfv\'en waves to propagate, their periods must be much less than the diffusion time $\tau_\eta=c^2/\eta$ (with $\eta\simeq1$ m$^2$.s$^{-1}$ the magnetic diffusivity). 
It is the case in the Earth's core, as measured by the Lundquist number $Lu=\tau_{\eta}/\tau_A\approx 10^{5}$. \\

\noindent When both rotation and magnetic field are present, the response to a planetary-scale perturbation by inertial waves is thus much faster than that by Alfv\'en waves. 
The predominance of the Coriolis force in the momentum budget has two important consequences. 
First, transient flows tend to be almost invariant along the rotation axis \citep{jault2008axial}, or quasi-geostrophic (QG).  
Second, for $Le\ll1$ there should exist no large length-scales QG Alfv\'en modes in Earth's core, but instead magneto-Coriolis (MC) modes \citep{hide1966free}. 
These eigen-modes in the presence of global rotation and an imposed field divide for $Le\ll1$ into slow and fast modes, whose periods are commensurate with respectively
\begin{equation}
\tau_{MC}^s\approx \tau_A^2(L/c)^2/\tau_{\Omega} \mbox{ and } \tau_{MC}^f\approx \tau_{\Omega}c/L\,, 
\end{equation}
with $L$ a typical length-scale \citep[for a review, see][]{finlay2010short}. 
Fast QG-MC modes are only weakly sensitive to the magnetic field and correspond to Rossby modes \citep[i.e. axially invariant inertial modes, see][]{zhang2001inertial}.  
At large length-scales their periods remain much smaller than $\tau_A$ (it spans approximately the range between 0.1 and 2 months for azimuthal wave numbers $m\le10$ and radial wave numbers $N\le5$). 
Because their ratio of magnetic to kinetic energy is weak their detection in magnetic data is very difficult (furthermore in the monthly period range the core signal is hindered by external sources -- see Figure~(\ref{fig:FFT})). 
Conversely, slow QG-MC (or ` magnetostrophic') modes are dominated by the magnetic energy, and their period reaches centuries to millennia. Thus, a priori, they cannot explain SA pulses (but see below).\\

\noindent Nevertheless one particular family of Alfv\'en modes exists. These `torsional' modes are organised as cylinders co-axial with the rotation axis (geostrophic cylinders). For these modes, the projection of the Coriolis force vanishes, so that the fluid response to a perturbation naturally comes from the Lorentz force \citep{braginski70}. 
Detected from geomagnetic field models at a period of 6-yrs \citep{Gillet:2010N}, their outward propagation across the fluid core in about 4 years constrains the intensity of the field in the bulk of the  core. 
However, their flow perturbation magnitude (of r.m.s. $\approx 0.3-0.5$ km.yr$^{-1}$) is not large enough to explain alone the observed SA pulses, and therefore non-axisymmetric flow contributions about 3 to 5 times larger are required \citep{gillet2015planetary,gillet2019reduced,Kloss_Finlay_2019}.\\

\noindent Several possibilities have been proposed to explain such intense non-zonal motions at inter-annual periods, among which are waves involving a stratified layer at the top of the core. 
The restoring force then comes from an interplay between the magnetic, Archimedes and Coriolis forces. 
Such `MAC' waves are appealing in the context of a heat flux at the CMB weaker than the adiabatic flux, a scenario made possible with a relatively high thermal conductivity of the core \citep{pozzo2012thermal,ohta2017thermal}, and motivated by seismological studies \citep{helffrich2010outer}. 
The role of buoyancy in the stable layer is governed by the steepness of the density profile $\rho'(r)=\rho(r)-\rho_0(r)$ below the CMB (with $\rho_0$ the reference density profile for stability, the adiabatic one in the Boussinesq approximation). 
It is measured by the Brunt-Va\"isala frequency 
\begin{equation}
N=\sqrt{-\frac{g}{\rho}\frac{d\rho'}{dr}}\,, 
\end{equation}
where $g(r)$ is the gravity acceleration. 
For reduced models of a hidden ocean at the top of the core, the control parameters are only the density jump $\Delta \rho$ across the layer, and the height $H$ of the stratified layer \citep{Braginsky93,braginsky1999dynamics}. 
By tuning $N$ (or $H$ and $\Delta \rho$), it is possible to match the inter-annual time-scales of interest for SA pulses \citep[e.g.,][]{chulliat2015fast,chiduran2020}. 
Note that the configuration of the background field has a significant influence on the spatial distribution of the waves \citep[see][]{knezek2018influence}.
In particular under some conditions MAC waves may be trapped in the equatorial belt \citep{Buffett_Matsui_2019}, as is the majority of SA pulses witnessed over the satellite era.  \\

\noindent Meanwhile, the existence of a stratified layer is still a debated issue from several aspects: seismology \citep{irving2018seismically}, geodynamo simulations \citep{gastine2020dynamo} and  high-pressure-temperature experiments \citep{konopkova2016}. 
Furthermore, rapid SV events have been found in geodynamo simulations, in the absence of any stratification, as the regime of parameters gets closer to Earth's conditions \citep{Aubert:2018}. 
There, they are interpreted as the signature of QG Alfv\'en waves carried by intense magnetic field lines \citep{aubert2019geomagnetic}. 
Such Alfv\'en waves can exist in spite of the Coriolis force because of their reduced extension in the cylindrical radial direction. 
If the Coriolis force determines their QG nature, it seems to only marginally affect their period, slightly above $\tau_A$ \citep{aubert2021}. Still the complexity of the background field within the dynamo simulations, with strong heterogeneities, renders more subtle to decipher the roles played by the Coriolis, Lorentz and diffusion terms along the wave propagation \citep{aubert2019approaching}.   
Even though SV field models are limited in spatial resolution, QG Alfv\'en waves might be detectable as they reach the equatorial area, because of the spherical geometry of the core that (i) focuses the wave energy and (ii) projects sharp gradients in the cylindrical radial axis onto smoother spatial variations on the spherical boundary.
The detection of such QG Alfv\'en waves in dynamo simulations is supported by the finding of \cite{gerick2020}: the periods of the slow QG-MC waves decrease to inter-annual values as the spatial complexity of its modes increases. 
These eigen-solutions furthermore focus the SV signals in the equatorial belt, and may explain the minimum of radial velocity at $\pm10^{\circ}$ in latitude recovered from magnetic data \citep{gillet2015planetary}. 
They may therefore be the elementary components needed to understand the SA derived from satellite data. 
The several families of waves discussed above, and their main characteristics regarding SA events, are summarised in Table \ref{tab: waves}.\\

\begin{table}
\caption{Waves in the inter-annual period range of interest for explaining SA changes, with their main physical ingredients.
$^a$ the period of long period inertial (or Rossby) waves increases towards $\tau_A$ with increasing spatial complexity \citep{zhang2001inertial}.
$^b$ detected from magnetic data \citep{Gillet:2010N}.
$^c$ from the eigen-mode study by \citep{gerick2020}.
$^d$ detected in numerical simulations \citep{aubert2019geomagnetic,aubert2021}.
$^{\dag}$ although essentially of magnetic nature, these waves still are inprinted by the Coriolis force (see text for details).
$^e$ for their focalization in the equatorial area, see \cite{Buffett_Matsui_2019}.
\label{tab: waves}}
\begin{tabular}{@{}|l|c|c|c|c|c|} \hline
\renewcommand{\baselinestretch}{0.85}
name & Coriolis & magnetic  & stratification & must involve short  & SA enhanced in  \\
& force & field & & wave lengths ? & the equatorial belt\\
 \hline
slow inertial waves$^{a}$ & yes & no & no & in the cylindrical& yes but carry too weak  \\
&&&&  radial direction, yes &magnetic energy\\
\hline
torsional Alfv\'en waves$^b$ & no & yes & no & not required for  & cannot alone explain \\
&&&& the gravest modes & the entire SA signal\\
\hline
QG-MC modes$^c$ &  yes &  yes &  no & in the cylindrical & yes \\
&&&&  radial direction, yes &\\
\hline
QG-Alfv\'en waves$^{d}$ &  yes$^{\dag}$ &  yes & no & yes & most often\\
\hline
MAC waves$^e$ &  yes &  yes & yes & yes (at least in the & possibly (depends on\\
&&&& radial direction) & the background field)\\
 \hline
\end{tabular}
\end{table}

\noindent One should keep in mind some caveats when interpreting the strong SA components in models of the core magnetic field. 
First, a significant part of the magnetic signal is likely associated with unresolved patterns, consequence of the advection process (equation~\ref{eq:de} \&~\ref{eq:dde}) that nonlinearly mixes wavelengths \citep[e.g.,][]{Pais:2008,gillet2019reduced}. 
This reduces the resolution of the reconstructed core dynamics.  
Furthermore, the quasi-periodic nature of observed SA pulses \citep{soloviev2017detection} remains questionable, because of the limited spatio-temporal resolution of geomagnetic field models \citep{gillet2019}, inherent in the difficult separation of magnetic sources \citep[see section~\ref{sec:gfm} and also][]{finlay2017challenges}. 
Indeed, when looking at rapid SV changes from magnetic field models, the most intense patterns necessarily show up at intermediate length-scales, as towards short wave-lengths fast variations are generally smoothed out (see Figure \ref{Fig:SA chart} and Section~\ref{sec:gfm}).\\

\noindent Finally, whatever the source of SA pulses, the core dynamics do generate a magnetic field with a temporal spectrum behaving approximately as $S(f)\propto f^{-4}$ at observatory sites, or equivalently (in the case of an insulating mantle) through series of geomagnetic Gauss coefficients at the core surface. 
This spectral property has been recovered in dynamo simulations, even when run at parameters far from Earth-like values, for the dipole moment  \citep{olson2012superchrons,buffett2015power} as well as for non-dipole coefficients \citep{bouligand2016frequency}. 
However, the range of frequencies relevant to interpret inter-annual changes must also be considered in comparison with the Alfv\'en time $\tau_A$ on the one side, and with the turn-over time $\tau_U=c/U$ on the other side. Here $U$ is the typical flow speed in the core, and $c$ the core radius, leading for $U\approx 20$ km.yr$^{-1}$ to $\tau_U\approx 150$ yr.
The spectral range in $f^{-4}$ for the Earth's core is found for frequencies in the range $\tau_A \le 1/f <\tau_U$. 
It is numerically very expensive to reach geophysical values of the Alfv\'en number $Al=\tau_A/\tau_U\approx10^{-2}$ in numerical geodynamo simulations \citep{schaeffer2017turbulent}. When $Al=O(1)$, all the processes at work in Earth's core at periods from a few years to a century are indistinguishable. 
It is only recently, by using specific parameterisation of turbulent processes, that values of $Al$ significantly lower than $O(1)$ have been achieved \citep{aubert2017spherical}. 
Only such extreme numerical experiments present a dynamics usable to replicate the physics of rapid SV changes. 
They show that torsional waves propagate predominantly outward, as recovered from magnetic records, and are constantly excited. 
Furthermore, as numerical geodynamo parameters get closer to Earth's conditions, jerk events become ubiquitous, and the range where the temporal spectrum for the SA is approximately flat (or equivalently that of the main field $\propto f^{-4}$) becomes wider \citep{Aubert:2018,aubert2021}. 

\section{Conclusion}
\label{sec: conclusion}

In this short overview we have described first the observational evidences of fast variations of the magnetic field generated in the liquid outer core of the Earth. These evidences directly derive from processing steps applied on magnetic observatory or satellite data. Both show episodes of fast variations of the geomagnetic field secular variation, mainly at mid- and low-latitudes. The application of modelling techniques either localised or global, shows that these fast variations are mainly due to strong, localised, spots of acceleration at the core mantle boundary. These are generally, but not always, located close to the equator under Indonesia or Central America, or in the northern hemisphere along the meridian 100$^\circ$E. \\

\noindent In order to understand these observations, one of the pressing questions that must be addressed is the temporal variability acceptable for core field models. An answer to this question will necessarily come from a combination of  theoretical approaches that provide the possible characteristics (periodicities, strength, location) of a signal generated in the core, and an observational approach that extracts core field signatures from the complex mix of signals observed at the Earth's surface or in its immediate environment. Regarding this latter approach, the data coverage provided by satellite missions is such that the main difficulty to answer this question is the characterisation and separation of the core field from sources of internal origins, as seen from satellite altitudes. The fields induced in the mantle by rapidly varying external fields and ionospheric contributions are the sources, internal to the satellite sampling region, most difficult to separate. \\

\noindent Current field models present time scales for the secular acceleration much shorter than previously thought, of the order 10 to 15 years for the largest wavelengths (spherical harmonic degrees 1 and 2), down to 6 to 7 years from spherical harmonic degrees 5 or 6 up. They present also a temporal variability that has not been yet studied, but certainly carries information on the core dynamics. However, these time scale values may still be over-estimated and hide part of the complexity of the underlying dynamics. In particular $\tau_{sa}$ remains much larger than the expected cut-off period (one year or less) where the magnetic core field variation would be smoothed out by the magnetic diffusion within the core or by electrical currents in the conductive mantle. Nevertheless, it is possible that the obtained $\tau_{sa}$ values are geophysically relevant since extrapolation of the current dynamo runs the closest to Earth's conditions suggest that time scales of $\approx$7 years are possible.\\ 

\noindent However, these dynamo simulations do not explore configurations where there is a stratification at the top of the core and alternative processes for generating inter-annual motions. We have tried here to give an overview of the possible transient responses that have been documented so far. 
It is not absolutely an exhaustive picture, as we ignore possible forcings from Earth nutation and precession that may be of importance as the Earth's core deviates from a perfect sphere \citep{le2015flows}. In a turbulent regime, these forcings may also lead to QG flows and trigger dynamics at inter-annual time-scales.\\

\noindent Currently, the scientific community has a limited knowledge on the three-dimensional distribution of the mantle conductivity. In addition we lack data at Earth's surface to better describe ionospheric fields, and we have difficulties to separate lithospheric or oceanic contributions from the core field. It follows that the most promising path to better describe the core field dynamics is to identify through theoretical studies their expected main characteristics (e.g., propagating waves, time scales) and try to identify their signals in observatory and satellite data sets. This can be done through a co-estimation with core surface flow models \citep[e.g.,][]{Barrois_etal_2018,barenzung2018modeling}, or more sophisticated assimilation approaches based either on dynamo simulations \citep[e.g.,][]{sanchez2019sequential} or on reduced QG equations \citep[see][]{Jault2015TOG,gerick2020,jackson2020plesio}, although these have not led yet to any application to geophysical data. 
Given the time scales involved in the core field processes and the core complex spatio-temporal dynamics, it is crucial to  continuously measure the magnetic field with satellites carrying magnetometers and in magnetic observatories. Some information content on the core dynamics might be obtained from other types of data, such as Earth's rotation, gravity or geodesy. The core variations signature in these data is certainly very weak, however their use in synergy with magnetic data will lastly improve our understanding of the liquid core dynamics.

\begin{acknowledgements}
We thank the anonymous reviewer for remarks that help us improving the manuscript. This work was partially supported by the CNES. It is based on observations with magnetometers embarked on Swarm mission. NG contribution has been also funded by ESA in the framework of EO Science for Society, through contract 4000127193/19/NL/IA (SWARM + 4D Deep Earth: Core). MM: the research leading to these results has received funding from the European Research Council (ERC) GRACEFUL Synergy Grant No 855677. MDH was supported by the European Research Council (ERC) under the European Union's Horizon 2020 research and innovation programme (grant agreement No. 772561) and also partly by Swarm DISC activities, funded by ESA contract no. 4000109587. This is the IPGP contribution number XXXX
\end{acknowledgements}

\bibliographystyle{./LaTeX_DL_468198_240419/spbasic}
\bibliography{./papier}

 \end{document}